\documentclass[10pt,letterpaper]{article}
\usepackage{graphicx}
\usepackage[margin=3cm]{geometry}
\usepackage{siunitx}
\DeclareSIUnit\decibelc{dBc}
\usepackage{amsmath,subfigure}
\allowdisplaybreaks
\newcommand{\e}{{\rm e}}
\begin{document}
\title{Self--generation of optical frequency comb in single section Quantum Dot Fabry-Perot lasers: a theoretical study}
\author{Paolo Bardella,$^1$ Lorenzo Luigi Columbo,$^{1,2}$ and Mariangela Gioannini$^{1^*}$}
\maketitle
\begin{center}
\noindent
$^1$ Dipartimento di Elettronica e Telecomunicazioni, Politecnico di Torino, \\ \noindent Corso Duca degli Abruzzi 24, Torino, IT-10129, Italy\\
$^2$ Consiglio Nazionale delle Ricerche, CNR-IFN, via Amendola 173, Bari, IT-70126, Italy\\ \noindent
*mariangela.gioannini@polito.it
\end{center}

\begin{abstract}
Optical Frequency Comb (OFC) generated by semiconductor lasers are currently widely used in the extremely timely field of high capacity optical interconnects and high precision spectroscopy. Very recently, several experimental evidences of spontaneous OFC generation have been reported in single section Quantum Dot (QD) lasers. 
Here we provide a physical understanding of these self-organization phenomena by simulating the multi-mode dynamics of a single section Fabry-Perot (FP) QD laser using a Time-Domain Traveling-Wave (TDTW) model that properly accounts for coherent radiation-matter interaction in the semiconductor active medium and includes the carrier grating generated by the optical standing wave pattern in the laser cavity. We show that the latter is the fundamental physical effect at the origin of the multi-mode spectrum appearing just above threshold. A self-mode-locking regime associated with the emission of OFC is achieved for higher bias currents and ascribed to nonlinear phase sensitive effects as Four Wave Mixing (FWM). Our results are in very good agreement with the experimental ones.
\end{abstract}

\section{Introduction}
An Optical Frequency Comb (OFC) is a light emission  characterized by equally spaced narrow optical lines with same intensity, low phase noise and low mode partition noise \cite{Hansch}; an OFC laser is therefore different from a multi-wavelength laser, because, in the latter, the lasing lines may be not phase locked and the coupling between them may lead to important mode partition noise.

In semiconductor lasers, OFC have been obtained with mode-locking techniques (either active mode-locking via RF electrical modulation of the laser gain/losses or passive massive mode-locking with reversely biased saturable absorber in laser cavity) or gain-switching.  A breakthrough has been represented by the experimental evidences of self-mode-locking in single section lasers  characterized by the absence of any active or passive modulation; self-generated OFC has been experimentally reported in different types of Fabry-Perot lasers based on Quantum Well (QW) \cite{Sato}, Quantum Dash (QDash) \cite{Panapakkam}, QD \cite{Liu} active materials and in Quantum Cascade Lasers (QCLs) \cite{faistreview}. Self-generated OFCs in \SI{1.3} {\micro\meter} and \SI{1.5} {\micro\meter} wavelength bands have been extensively investigated in QD and QDash FP laser under CW injection \cite{Gosset,Merghem1, LuCombSopraSoglia, Liu, Lu, Merghem,  Rosales1, Rosales2, Rosales3}; pulse trains have been reported directly at the laser output \cite{Gosset, Merghem1, LuCombSopraSoglia, Liu, Lu} or after compensation of the group velocity dispersion trough, e.g., a dispersive optical fiber of suitable length \cite{Merghem, Rosales1, Rosales2, Rosales3}. Despite the observation of pulses is an unambiguous fingerprint of the OFC formation, we evidence that OFC lasers generate lasing lines that are in general phase locked, even when there is no regular phase relation between the lines \cite{faistreview}. For many applications, as those discussed below, the phase locking condition (leading to narrow RF beat-note and narrow optical linewidths) is the sole requirement and regular output pulses are not strictly needed.

These very simple OFC sources have attracted an impressive interest  in the rapidly growing field of high-data rates DWDM optical interconnection where the OFC laser diode feeds the silicon photonics optical modulators  to realize a compact and low cost transmitter \cite{Delfyett, Chen, Eiselt}. Another interesting application is in the field of photonic generation of sub-THz signals via the mixing of the comb lines, spaced of hundreds of GHz, on a fast photodetector \cite{Koenig}. QCL in the mid-infrared range are used for high precision and high speed spectroscopy based on dual comb techniques \cite{link2017, faistreview}.

The understanding of the fundamental physical effects at the origin of self-generation of OFC in FP semiconductor lasers is still matter of present research. Multi-wavelength lasing  close to threshold is generally ascribed to  fast carrier gratings with sub-wavelength spatial variation which is generated by the optical field standing wave pattern in the FP cavity (see e.g. \cite{Moloney1}). This longitudinal modulation of the carriers (spatial hole burning, SHB) destabilizes the single longitudinal mode emission and favors  the onset of several longitudinal cavity modes with different spatial configuration. SHB is rather week in QW lasers because the carrier diffusion tends washing out the carrier gratings while the cross-coupling between longitudinal cavity modes is responsible for the high mode partition noise \cite{Lenstra,Giudici}.

In low dimensional active materials as QDs and QDashes, the carriers are trapped in the quantum box and cannot diffuse, therefore the SHB is much stronger than in the QW counterpart. From this viewpoint,  QD and QDash laser diodes are  similar to QCLs, where carrier diffusion time is longer than the typical carriers life time \cite{FaistBook}. Despite SHB alone can explain the multi-wavelength lasing, it is not sufficient to enable the phase-locking among the lasing spectral lines \cite{Gordon, Boiko,Jirauschek}.

Recent theoretical analysis carried on for the QCL case (see e.g. \cite{Villares2015}) have attributed the phase-locking and the consequent OFC generation to nonlinear phase sensitive effects, like FWM, which are inherent to the saturable character of radiation-matter interaction in SCLs and impose phase-matching conditions for sufficiently high intra cavity field (ie: high CW bias current) and when the dispersion is limited. For standard QW laser diodes, the rigorous description of the role played by the FWM effect in the phase locking of three cavity longitudinal modes has been presented in \cite{Landais}. In \cite{Landais}, the Authors also identified  the operating conditions where the FWM is able to compensate the  non uniform frequency spacing among the cold cavity modes, thus leading to self-phase-locking.

In our previous work \cite{Gioannini}, we have studied the formation of OFC in  FP single section QD lasers using a Time Domain Traveling Wave model. In that work, starting from the assumption that the carrier gratings originated by the standing wave pattern could lead to a compression of the  gain at the wavelength of the lasing mode \cite{Mecozzi,Serrano}, we have introduced an empirical self- gain compression factor ($\epsilon$-parameter) of the QD  material gain and we have demonstrated that it led to a multi-wavelength emission and, in some cases, to spontaneous OFCs generation.

Here, upgrading that model \cite{Rossetti, Gioannini}, we present a TDTW model that eliminates the need of an empirical $\epsilon$-parameter and includes in a rigorous way the effect of SHB. With rigorous numerical simulations and analysis of the results, we demonstrate that the SHB is, in QD systems, the fundamental physical effect at the origin of the multi-wavelength lasing spectra always observed experimentally at any current  just above threshold. Moreover, thanks to our model, we can identify in this work intervals of the bias current where, together with multi-wavelength emission, we have also an actual OFC thanks to FWM which is responsible for mutual injection locking of the cavity modes.

The paper is organized as follows: in section 2, we describe the TDTW model used for simulating the multi-mode dynamics of the QD laser and, in section 3, we present the results of numerical simulations for a standard QD  single section FP laser. In section 4 we compare our observations on comb self-generation with the experimental evidences reported in the existing literature and we provide a physical interpretation of the self-phase locking. Finally we draw our conclusions in Section 5.

\section{QD laser TDTW model in presence of carrier grating}
We consider a single section Quantum-Dots-in-a-Well (DWELL) InAs/GaAs laser emitting from the ground state (GS) around \SI{1258}{\nano\meter}. The length of the FP cavity ($L$) is a few hundred microns. We sketch in Fig.\,(1) the typical device configuration (a) and the electron dynamics (b) as considered in our model, whereas the main material and device parameters are summarized in Table \ref{Tableparam}.

The SHB is introduced in the model adopting a multiple scale approach where a slowly varying envelope approximation (SVEA) is adopted for describing the spatio-temporal evolution of the optical electric field; this approximation allows discretizing the cavity length with spatial steps of few microns including, nevertheless, the effects of the fast standing wave pattern on carrier dynamics \cite{Gordon, Moloney1, Javaloyes, Balle1}.

In the SVEA, the optical electric field of the fundamental TE mode is written as: $$\hat{E}(x,y,z,t)=\phi(x,y)\left\{\left[E(z,t)^{+}\exp{(-jk_{0}z+j\omega_{0}t)}+E(z,t)^{-}\exp{(+jk_{0}z+j\omega_{0}t)}\right]+c.c.\right \}$$ where $\phi(x,y)$ is the transverse mode profile that we assume independent on $z$, $\omega_{0}$ is our reference angular frequency and $k_0$ is the propagation constant of the cold cavity TE mode at $\omega_{0}$. The terms $E(z,t)^{\pm}$ represent the slowly varying spatio-temporal variation of the forward and backward components of the electric field that obey the following equations (\cite{Moloney1}):
\begin{equation}
\frac{1}{v_{g}}\frac{\partial E^{\pm}}{\partial t} \pm \frac{\partial E^{\pm}}{\partial z}= -\frac{\alpha_{wg}}{2}E^{\pm}-j\frac{\omega_{0}}{2c\eta \epsilon_{0}}\Gamma_{xy} <P\exp{(\pm jk_{0}z)} >+S_{sp}^{\pm}
     \label{field1}
\end{equation}
In the FP configuration the forward and backward field envelops satisfy the boundary conditions:
\begin{equation*}
E^{+}(0,t)=r_{R}E^{-}(0,t), \, E^{-}(L,t)=r_{F}E^{+}(L,t)
\end{equation*}
where $r_{R,F}$ are the reflectivities at the rear ($z=0$) and front ($z=L$) facets, respectively.\\
\begin{figure}[ht]
	\centering
	\subfigure[]
	{\includegraphics[width=4.3cm]{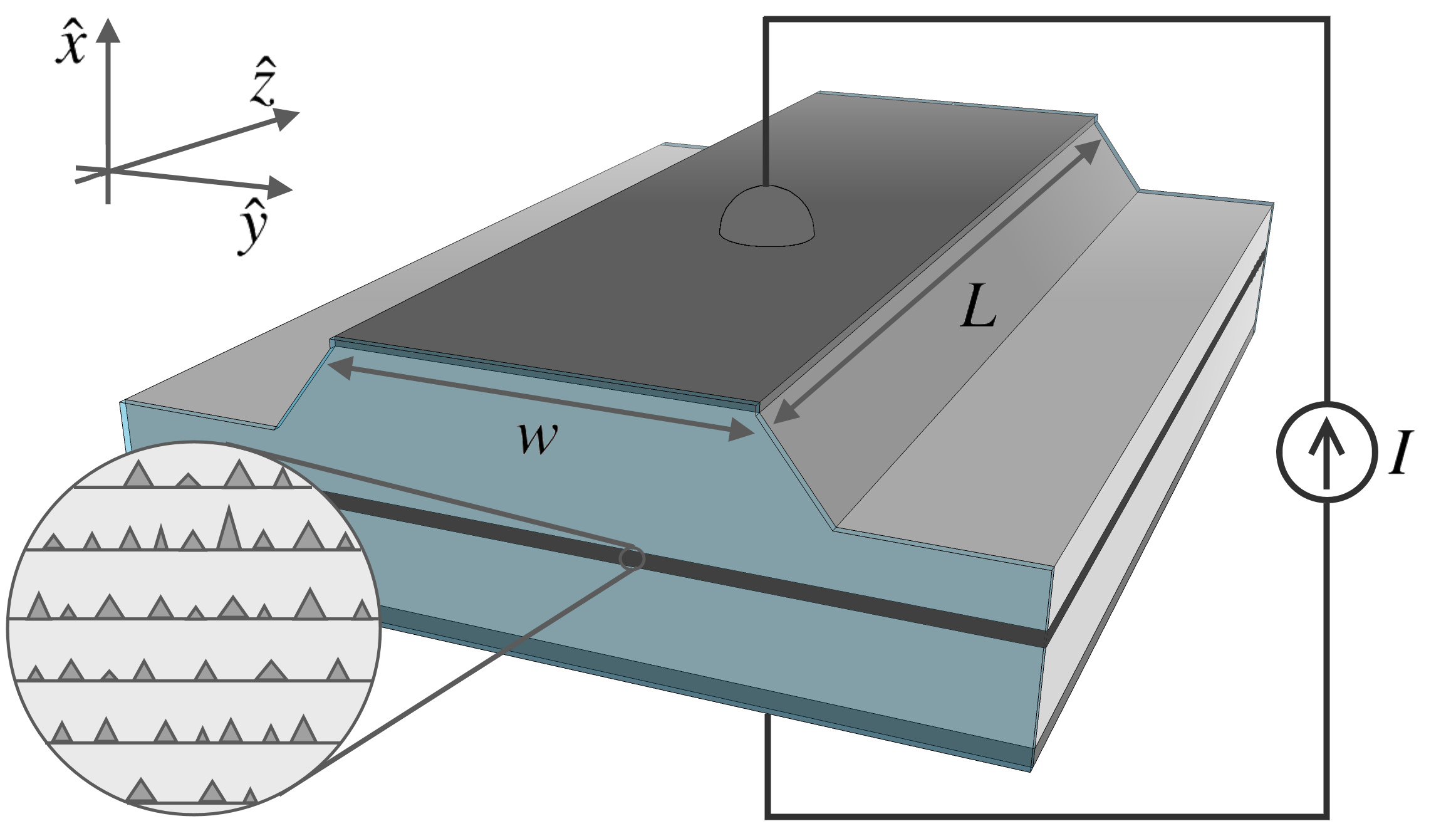}}
	\hspace{0mm}
	\subfigure[]
	{\includegraphics[width=4.3cm]{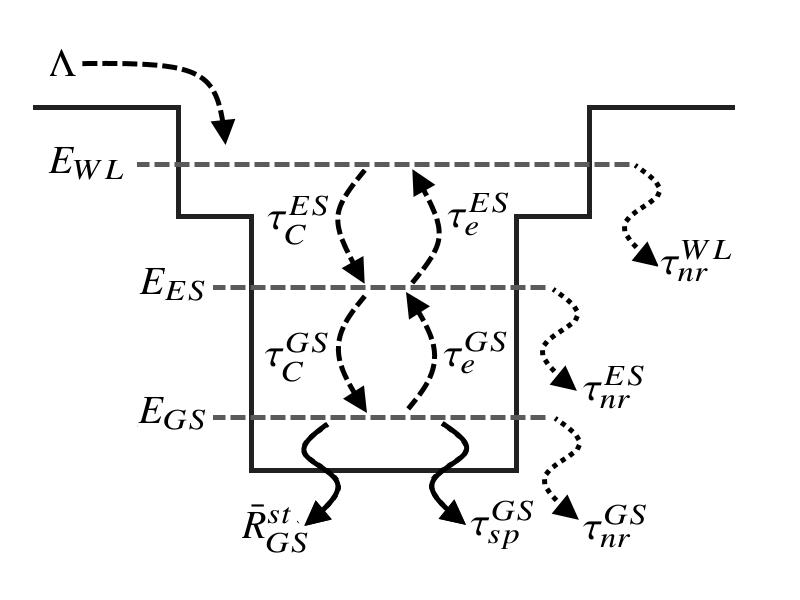}}
        \label{fig0}
        \caption{Schematic of the simulated single section QD laser (a) and schematic of the electron dynamics as considered in our model (b).}
\end{figure}

The variable $P(z,t)$ is the complex macroscopic polarization slowly varying in time (having extracted the component $\exp(j\omega_{0}t)$), but still including the fast spatial variation in the $z$-direction. 
The symbol $<\cdot>$ denotes in (\ref{field1}) a spatial average over many wavelengths along the $z$-direction; $v_{g}$ is the group velocity of optical mode in the waveguide, $\Gamma_{xy}$ is the transverse optical confinement factor in the active region, $\alpha_{wg}$ are the waveguide losses per unit length and $S_{sp}^{\pm}$ is the noise source associated with the process of spontaneous emission.\\
The relation between the macroscopic polarization $P$ and the microscopic polarizations $p_{m}(z,t)$ due to the GS and Excited State (ES) levels of each QD of the ensemble is:
$$
P(z,t)=\frac{N_{D}}{h_{QD}} \sum_{m=GS, ES} \mu_{m}d_{m}^{*}p_{m}(z,t)
$$
where $N_{D}$ is the number of QDs per unit area, $h_{QD}$ is the thickness of a single QD layer,  $\mu_{m}$ is the degeneracy of the $m$-th level (with $m=GS, ES$) and $d_{m}$ is the dipole matrix element associated to the optical transition from level $m$-th.\\
The variables $p_{m}(z,t)$, with $m=GS,ES$, satisfy the following {\it two-level like} dynamical equations:
\begin{equation}
\frac{\partial p_{m}(z,t)}{\partial t}=[j(\omega_{m}-\omega_{0})-\Gamma]p_{m}+j\frac{d_{m}}{\hbar}(2\rho_{m}-1)\left[E^{+}\exp{(-jk_{0}z)}+E^{-}\exp{(jk_{0}z)}\right]\label{Pfast1}
\end{equation}
where $\hbar \omega_m$ is the energy associated to the GS or ES transition and $1/\Gamma$ is the dephasing time which in our model is set to fit the gain dispersion of the QD ensemble. The variables $\rho_{m}(z,t)$ represent the electron occupation probability in the $m$-th level and keep the fast spatial variation of the carrier density due to the SHB; we assume that the hole dynamics will follow the electron dynamics (excitonic approximation).\\
We adopt in the following the scalings: $$E^{\pm} \longrightarrow E^{\pm} \sqrt{\frac{c \eta \epsilon_{0}wh_{QD}\eta_{L}}{2 \Gamma_{xy}}}, \, P^{\pm} \longrightarrow P^{\pm} \sqrt{\frac{c \eta \epsilon_{0}wh_{QD}\eta_{L}}{2 \Gamma_{xy}}}$$
where $w$ is the ridge area and $\eta_{L}$ is the number of QD layers, such that the quantity $|E^{\pm}(z,t)|^{2}$ represents the total power across the device transverse section.\\

\sisetup{list-units=single,list-final-separator = {, }, list-pair-separator=
	{, }} 

\begin{table*}[tb] \caption{Main Model Parameters}
	\small
	\label{tab:MainModelParameters} \renewcommand{\arraystretch}{1.1} 
	\begin{center}
		\begin{minipage}[c]{\textwidth}
			\centering
			
			\begin{tabular}{ccc} \hline \hline Symbol & Description & Values\\ \hline
				\multicolumn{3}{c}{Material parameters}\\ \hline
				$\eta$&Effective refractive
				index &\SI{3.34}{}\\ $n_L$&Number of QD layers&\SI{15}{}\\ $N_D$&QD surface
				density&\SI{2.7e10}{\per\centi\metre\squared}\\
				$\mu_{m}$&Confined states degeneracy $m = ES, GS$&\SIlist{4;2}{}\\
				$1/\Gamma$&Dipole dephasing time&\SI{15}{\femto\second}\\
				$d_{GS}$&Dipole matrix element for GS&\SI{0.72}{\electronvolt \nano\meter}\\
				$\tau_{C}^{m}$& Electron capture times $m = GS, ES$&\SIlist[parse-numbers=false]{6;1}{\pico\second}\\
				$\tau_{e}^{m}$& Electron escape times $m = GS, ES$&\SIlist[parse-numbers=false]{10.8;4.1}{\pico\second}\\
				$\tau_{nr}^{m}$& Electron non-radiative decay times $m = GS, ES, WL$&\SIlist[parse-numbers=false]{1,1,1}{\nano\second}\\
				$\tau_{SP}^{GS}$& Electron spontaneous emission time for GS&\SI{2}{\nano\second}\\
				$\alpha_{wg}$&Intrinsic  waveguide losses&\SI{5}{\per\centi\meter}\\ \hline
				\multicolumn{3}{c}{Device parameters}\\ \hline 
				$w$&Ridge width&\SI{5}{\micro\metre}\\ 
				$h_{QD}$&QD layer thickness&\SI{5}{\nano\metre} \\ 
				$L$&Device  length&\SI{250}{\micro\metre}\\ $\Gamma_{xy}$& Transverse optical confinement factor&\SI{12}{\percent}\\ $r_R^{2}, r_F^{2}$& Terminal facets
				reflectivity &\SI{60}{\percent}\\ \hline
			\end{tabular}
			\label{Tableparam}		
			
		\end{minipage}
	\end{center}
\end{table*}

\vspace{0.3 cm}

The dynamics of the carriers in the GS, ES and in the wetting layer (WL) is described via a set of rate equations that read \cite{Rossetti}:
\begin{eqnarray}
\frac{\partial \rho_{GS}(z,t)}{\partial t}&=& -\frac{\rho_{GS}}{\tau_{nr}^{GS}} -\frac{\rho_{GS}^{2}}{\tau_{sp}^{GS}} -\frac{\rho_{ GS}(1-\rho_{ ES})}{\tau_{e}^{GS}}+\frac{\mu_{ES} }{\mu_{GS}}\frac{\rho_{ ES}(1-\rho_{GS})}{\tau_{C}^{GS}}    \nonumber        \\
&\phantom{=}& -\frac{\bar{R}_{ GS}^{st}}{N_{D}\mu_{GS}\eta_{L}} \label{pop1fasta1} \\
\frac{\partial \rho_{ES}(z,t)}{\partial t}&=& -\frac{\rho_ ES}{\tau_{nr}^{ES}}
+\frac{D_{WL}}{\mu_{ES}N_d}\frac{\rho_{WL}(1-\rho_{ES})}{\tau_{C}^{ES}}-\frac{\rho_{ ES}(1-\rho_{WL})}{\tau_{e}^{ES}} \nonumber\\
&+&\frac{\mu_{GS} }{\mu_{ES}}\frac{\rho_{GS}(1-\rho_{ ES})}{\tau_{e}^{GS}}-\frac{\rho_{ES}(1-\rho_{ GS})}{\tau_{C}^{GS}}
\label{pop2fastb}\\
\frac{\partial \rho_{WL}(z,t)}{\partial t}&=& \Lambda-\frac{\rho_{WL}}{\tau_{nr}^{WL}} +\left[-\frac{\rho_{WL}}{\tau_{C}^{ES}}(1-\rho_{ES})+\frac{\mu_{ES}N_{D}}{D_{WL}}\frac{\rho_ES}{\tau_{e}^{ES}}(1-\rho_{WL})\right] \label{pop2fastc}
\end{eqnarray}
where the stimulated emission rate per unit area has the following expression: $$\bar{R}_{GS}^{st}=\frac{-j \,\beta_{GS}}{2} \times \left[\left(E^{+*}p_{GS}-E^{-}p_{GS}^{*}\right)\exp{(jk_{0}z)}+\left(E^{-*}p_{GS}-E^{+}p_{GS}^{*}\right)\exp{(-jk_{0}z)}\right]$$
with $$ \beta_{GS}=\frac{2\Gamma g_{0,GS}}{w \, \omega_{GS}d_{GS}}, \quad g_{0, GS}=\frac{\omega_{0} N_{D} \Gamma_{xy} \mu_{GS}d_{GS}^{2}}{2c \eta \epsilon_{0} \hbar \Gamma h_{QD}} $$
where $\eta$ is the effective refractive index.
In Eqs. (\ref{pop2fastb})-(\ref{pop2fastc}), $\rho_{WL}$ is the occupation probability of the WL which is represented as a single energy level with degeneracy per unit area $D_{WL}$,  
$\eta_{L}$ is the number of QD layers, $\Lambda= I\,\eta_{i}/(e\, D_{WL} \eta_{L} \, w \,L)$ is the carrier injection rate in the WL, $\eta_{i}$ is the internal quantum efficiency, and $I$ the pump current.
Referring to Fig.\,\ref{fig0}, the characteristic times $\tau_{C,e}^{GS}$ represent the capture ($C$) and the escape time (e) between GS and ES. Similarly, the capture/escape processes between WL and ES is quantified by the time constants $\tau_{C,e}^{ES}$.  The carrier lifetime due to spontaneous and non-radiative recombinations in the states is quantified by $\tau_{sp,nr}^{m}$ with $m=GS, ES, WL$. Since we study here laser emission from the GS only, we will assume that the photon emission process occurs only in the GS.\\

To  couple the rate equations (\ref{pop1fasta1})-(\ref{pop2fastc}) with the slowly varying traveling wave equation of the electric field in Eq. (\ref{field1}), we expand in Fourier series the spatial variation at the wavelength scale of both the polarizations and the occupation probabilities:
\begin{subequations}
\begin{eqnarray*}
p_{GS}(z,t)&=&p_{GS}^{+}(z,t)\exp{(-jk_{0}z)}+p_{GS}^{-}(z,t)\exp{(jk_{0}z)}\label{FE1}\\ 
&=& \exp{(-jk_{0}z)} \sum_{n=0}^{+\infty}p_{n, GS}^{+}(z,t)\exp{(-2n jk_{0}z)}\\
&\phantom{=}&+\exp{(jk_{0}z)} \sum_{n=0}^{+\infty}p_{n, GS}^{-}(z,t)\exp{(2n jk_{0}z)} \\
\rho_{m}(z,t)&=&\rho_{0, m}(z,t)+\sum_{n=1}^{+\infty} \left[\rho_{n, m}(z,t)\exp{(-2n jk_{0}z)}+ c.c.\right]\\
\label{FE2}
\end{eqnarray*}
where $p_{n, GS}^{\pm}(z,t)$, $\rho_{n, m}(z,t)$ with $m = GS, ES, WL$ and $n=0,1, 2...$ are slowly varying functions of $z$. Inserting these expression in Eqs. (\ref{field1})-(\ref{pop2fastc}) and neglecting the terms with spatial frequency higher than $2k_{0}$, we finally get the following system of differential equations:
\begin{eqnarray}
\frac{1}{v_{g}}\frac{\partial E^{\pm}}{\partial t} \pm \frac{\partial E^{\pm}}{\partial z}&=&-\frac{\alpha_{wg}}{2}E^{\pm}-j\frac{\omega_{0}}{2c\eta \epsilon_{0}} \frac{\Gamma_{xy}\mu_{GS}d_{GS}N_{D}}{h_{QD}}p_{0, GS}^{\pm}+S_{sp}^{\pm} \label{fieldfasta}\\
\frac{\partial p_{0, GS}^{\pm}(z,t)}{\partial t}&=&[j(\omega_{GS}-\omega_{0})-\Gamma]p_{0, GS}^{\pm}+j\frac{d_{GS}}{\hbar}\left[(2\rho_{0, GS}-1)E^{\pm}+2\rho_{GS}^{\pm}E^{\mp}\right]\label{Pfast1a}\\
\frac{\partial \rho_{0,GS}(z,t)}{\partial t}&=& -\frac{\rho_{0,GS}}{\tau_{nr}^{GS}}- \frac{\rho_{0, GS}^{2}}{\tau_{sp}^{GS}} -\frac{\rho_{0,GS}(1-\rho_{0, ES})}{\tau_{e}^{GS}}+\frac{\mu_{ES} }{\mu_{GS}}\frac{\rho_{0, ES}(1-\rho_{0,GS})}{\tau_{C}^{GS}} \nonumber \\
&\phantom{=}&-\frac{\beta_{GS}}{\mu_{GS}N_{D}\eta_{L}} Im\left(E^{+*}p_{0, GS}^{+}+E^{-*}p_{0, GS}^{-} \right)  \label{pop1fasta}\\
\frac{\partial \rho_{GS}^{+}(z,t)}{\partial t}&=&  -\frac{\rho_{GS}^{+}}{\tau_{nr}^{GS}} -\frac{\rho_{GS}^{+}(1-\rho_{0, ES})}{\tau_{e}^{GS}}+\frac{\rho_{0,GS}\rho_{ES}^{+}}{\tau_{e}^{GS}}
-\frac{\mu_{ES} }{\mu_{GS}}\frac{\rho_{0,ES}\rho_{GS}^{+}}{\tau_{C}^{GS}}  \nonumber \\
&\phantom{=}& +\frac{\mu_{ES} }{\mu_{GS}}\frac{\rho_{ES}^{+}(1-\rho_{0,GS})}{\tau_{C}^{GS}}-\frac{\rho_{0, GS}\rho_{GS}^{+}}{\tau_{sp}^{GS}}\nonumber\\
&\phantom{=}& +j\frac{\beta_{GS}}{2\mu_{GS}N_{D}\eta_{L}}\left(E^{-*}p_{0,GS}^{+}-E^{+}p_{0,GS}^{-*} \right)  \label{pop2fasta}\\
\frac{\partial \rho_{0,ES}(z,t)}{\partial t}&=& -\frac{\rho_{0,ES}}{\tau_{nr}^{ES}}
+\frac{D_{WL}}{\mu_{ES}N_d}\frac{\rho_{0,WL}(1-\rho_{0,ES})}{\tau_{C}^{ES}}-\frac{\rho_{0,ES}(1-\rho_{0,WL})}{\tau_{e}^{ES}} \nonumber\\
&\phantom{=}&+\frac{\mu_{GS} }{\mu_{ES}}\frac{\rho_{0,GS}(1-\rho_{0, ES})}{\tau_{e}^{GS}}-\frac{\rho_{0,ES}(1-\rho_{0,GS})}{\tau_{C}^{GS}}
\label{pop2fasta2}\\
\frac{\partial \rho_{ES}^{+}(z,t)}{\partial t}&=&
    -\frac{\rho_{ES}^{+}}{\tau_{nr}^{ES}}
    +\frac{\mu_{GS} }{\mu_{ES}}\frac{\rho_{GS}^{+}(1-\rho_{0,ES})}{\tau_{e}^{GS}}
    -\frac{\mu_{GS} }{\mu_{ES}}\frac{\rho_{0,GS}\rho_{ES}^{+}}{\tau_{e}^{GS}}
    +\frac{\rho_{0,ES}\rho_{GS}^{+}}{\tau_{C}^{GS}}    \nonumber\\
    &\phantom{=}&
     -\frac{\rho_{ES}^{+}(1-\rho_{0, GS})}{\tau_{C}^{GS}}
    -\frac{D_{WL}}{\mu_{ES}N_{D}}
        \left[
        \frac{\rho_{0,WL}\rho_{ES}^{+}}{\tau_{
        C}^{ES}}
        -\frac{\rho_{WL}^{+}(1-\rho_{0,ES})}{\tau_{C}^{ES}}
        \right]\nonumber\\
        &\phantom{=}&
    +\frac{\rho_{0,ES}\rho_{WL}^{+}}{\tau_{e}^{ES}}    
    -\frac{\rho_{ES}^{+}(1-\rho_{0,WL})}{\tau_{e}^{ES}} \label{popESfasta} \\
\frac{\partial \rho_{0,WL}(z,t)}{\partial t}&=& \Lambda-\frac{\rho_{0,WL}}{\tau_{nr}^{WL}} -\frac{\rho_{0,WL}}{\tau_{C}^{ES}}(1-\rho_{0,ES})+\frac{\mu_{GS}N_{D}}{D_{WL}}\frac{\rho_{0,ES}}{\tau_{e}^{ES}}(1-\rho_{0,WL}) \label{pop2fasta3}\\
\frac{\partial \rho_{WL}^{+}(z,t)}{\partial t}&=&-\frac{\rho_{WL}^{+}}{\tau_{nr}^{WL}} +\frac{\rho_{0,WL}}{\tau_{C}^{ES}}\rho_{ES}^{+}-\frac{\rho_{WL}^{+}}{\tau_{C}^{ES}}(1-\rho_{0,ES})  \nonumber \\
&\phantom{=}& -\frac{\mu_{ES}N_{D}}{D_{WL}}
 \left(
        \frac{\rho_{0, ES}\rho_{WL}^{+}}{\tau_{e}^{ES}}
        -\frac{\rho_{ES}^{+}(1-\rho_{0,WL})}{\tau_{e}^{ES}}
        \right)  \label{pop3fasta}
\end{eqnarray}
\end{subequations}
where we used the notation:
$$\rho_{m}(z,t)\simeq \rho_{0,m}(z,t)+\rho_{m}^{+}(z,t)\exp{(-2k_{0}z)}+\rho_{m}^{-}(z,t)\exp{(2k_{0}z),}$$
with $\rho^{-}_{m}(z,t)^{*} = \rho_{m}^{+}(z,t)$ and $m = GS, ES, WL$.\\

In the dynamical equations (\ref{fieldfasta})-(\ref{pop3fasta}), we do not include the effect of higher order terms such as the coherence grating ($p_{1, GS}$). These terms would leave qualitatively unchanged the system multi-mode behavior \cite{Boiko}.

We observe that, if we neglect the carrier grating at the sub-wavelength scale by forcing ($\rho_{m}^{+}=0$ with $m=GS, ES, WL$), the model reduces to the TDTW model presented in \cite{Rossetti, Gioannini}.

 \section{Results of dynamical simulations}

In this Section we report numerical results on the multi-wavelength QD laser dynamics obtained by integration of the PDEs (Eqs. (\ref{fieldfasta})-(\ref{pop3fasta})) using an optimized finite difference algorithm described in details in the Appendix. Since the temporal variations of the carrier densities occur in the time scale from hundreds of femtoseconds to few picoseconds (which are slower than the polarization dephasing time $1/\Gamma$ set to \SI{15}{\femto\second}) we can solve the QD active material frequency dependent optical response as an infinite impulse response filter as described in the Appendix.\\
We discuss first the key role of the fast carrier grating in generating the multi-wavelength lasing regime at all currents above threshold and then we focus on the conditions for having the phase-locking of the lasing lines. In the following analysis, we start from the total field at the laser output facet $E_{out}(t)=\sqrt{1-r_{R}^{2}}E(L,t)$ that allows to calculate the total power emitted by the QD laser as $P(t)=|E_{out}(t)|^{2}$.
By numerically filtering $E_{out}(t)$ with Hanning windows of spectral width $W_h=0.8 FSR$ centered at the relevant peaks of the optical spectrum, we are able to isolate the temporal evolution of the field of each $q$-th lasing line $E_q(t)$ and then retrieve the power $P_q(t)=|E_{q}(t)|^{2}$ and the instantaneous phase $\phi_q(t)=\angle {E_{q}(t)}$ .\\
In order to characterize the multi-mode dynamics in the different simulated regimes, we introduce convenient parameters
that will measure the intensity noise and the amount of phase locking. We will use the notation $<\cdot>$ to indicate the temporal average, the symbols $\mu_q$ and $\sigma_q$ to indicate mean and standard deviation in time for each line, and $M$ to indicate the average over all the $N$ optical lines within the \SI{-10}{\decibel} optical bandwidth. These parameters are:

\begin{itemize}
\item The average output power  $\mu_{Pq}=<P_q(t)>$, the intensity noise $\delta P_q(t)=P_q(t)-\mu_{Pq}$ and the r.m.s of the intensity noise $\sigma_{\delta P_q}=\sqrt {<\left(\delta P_{q}\right )^2> }$ of the $q$-th line;
we will use the average value over the $N$ lines $M_{\sigma_ {\delta P}}=\frac{1}{N} \Sigma_{q=1}^N{\sigma_{\delta P_q}}$ as a measure of the average intensity noise.\\
\item The average optical frequency of the $q$-th line $\mu_{fq}=<\frac{1}{2\pi}\frac{\textrm{d}}{\textrm{d}t}\phi_q(t)>$ and the associated  phase fluctuations $\Phi_q(t)=\phi_q(t)-2 \pi\mu_{fq}t$.\\
\item The differential phase fluctuation between two adjacent lines $\Delta\Phi_{q}(t)=\Phi_{q+1}(t)-\Phi_{q}(t)$
 and its average value $\mu_{\Delta\Phi q}=<\Delta\Phi_{q}(t)>$; the corresponding standard deviation $\sigma_{\Delta\Phi_q}=\sqrt {<\left( \Delta\Phi_{q}(t)-\mu_{\Delta\Phi q}\right)^2>}$ is a measurement of the differential phase noise, whereas $M_{\sigma\Delta\Phi}=\frac{1}{N}  \Sigma_{q=1}^N{\sigma_{\Delta\Phi_q}}$ is the average differential phase noise.\\
\item The frequency separation between couples of adjacent lines $\Delta\mu_{fq}=\mu_{fq+1}-\mu_{fq}$;  $M_{\Delta\mu f}$ represents the average value for the modes within the \SI{-10}{\decibel} optical bandwidth. Our model does not include the waveguide dispersion, but, because of the QD active material dispersion, the longitudinal mode separation under current injection might be different from the cold FSR and it might change among  couple of modes  and with current.
\end{itemize}
Respect to parameters defined above, an ideal OFC would have for any $q$-th line
\begin{itemize}
\item $\sigma_{\delta P_q}$$=$$0$ indicating that the mode partition noise is completely suppressed
\item $\sigma_{\Delta \Phi_q}$ $=$ $0$ as consequence of the phase-locking, i.e. no phase fluctuation and therefore extremely narrow optical linewidth
\item the same value of $\Delta \mu_{f q}$ for any $q$-th line because, by definition, OFC requires equally spaced optical lines.
\end{itemize}
In order to better compare our results with the experimental evidences so far reported in the literature (see e.g. \cite{Muller}), we will also provide in this Section a calculation of more conventional (and experimentally more easily accessible) quantifiers  of the degree of coherence in the system. These are:

\begin{itemize}
 \item The Relative Intensity Noise (RIN) spectrum calculated for the total output power and the power of each line as, respectively:
 $$RIN(f)=|{\cal F}\{P(t)-<P(t)>\}|^{2}/<P(t)>^{2}, \quad RIN_{q}(f)=|{\cal F}\{P_{q}(t)-\mu_{Pq}\}|^{2}/\mu_{Pq}^{2}$$
 where ${\cal F}\{\cdot\}$ denotes the Fourier transform operator.
\item The integrated RIN in the electrical bandwidth $B$ ranging from \SI{10}{\mega\hertz} to \SI{21}{\giga\hertz} calculated for the total output power and for each line as, respectively:
$$iRIN=\frac{\int_{B}RIN(f)df}{B}, \quad iRIN_{q}=\frac{\int_{B}RIN_{q}(f)df}{B};$$ the quantity $M_{iRIN}$ represents the average value of $iRIN_q$, calculated for each $q$ line in the \SI{-10}{\decibel} optical bandwidth.
\item Beat note Radio Frequency (RF) spectrum.
\end{itemize}

\subsection{Role of SHB in the multi-mode dynamics}

In the following, we report simulation results for a \SI{250}{\micro\meter} long FP laser with cold cavity FSR of \SI{178}{\giga\hertz}.

\begin{figure}
	\centering
	\subfigure[]
	{\includegraphics[height=4.7cm]{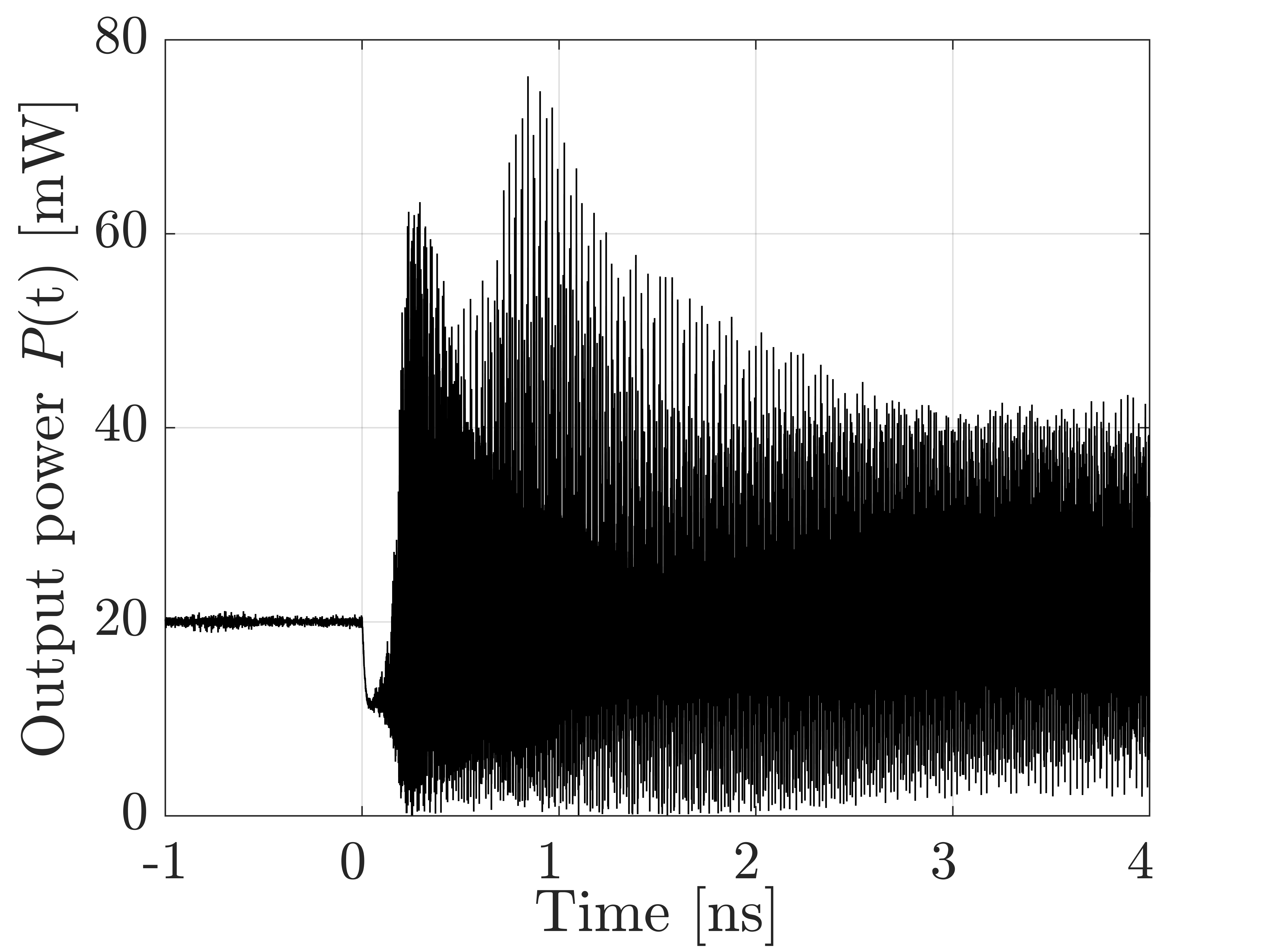}}
	\hspace{0mm}\\
	\subfigure[]
	{\includegraphics[height=4.7cm]{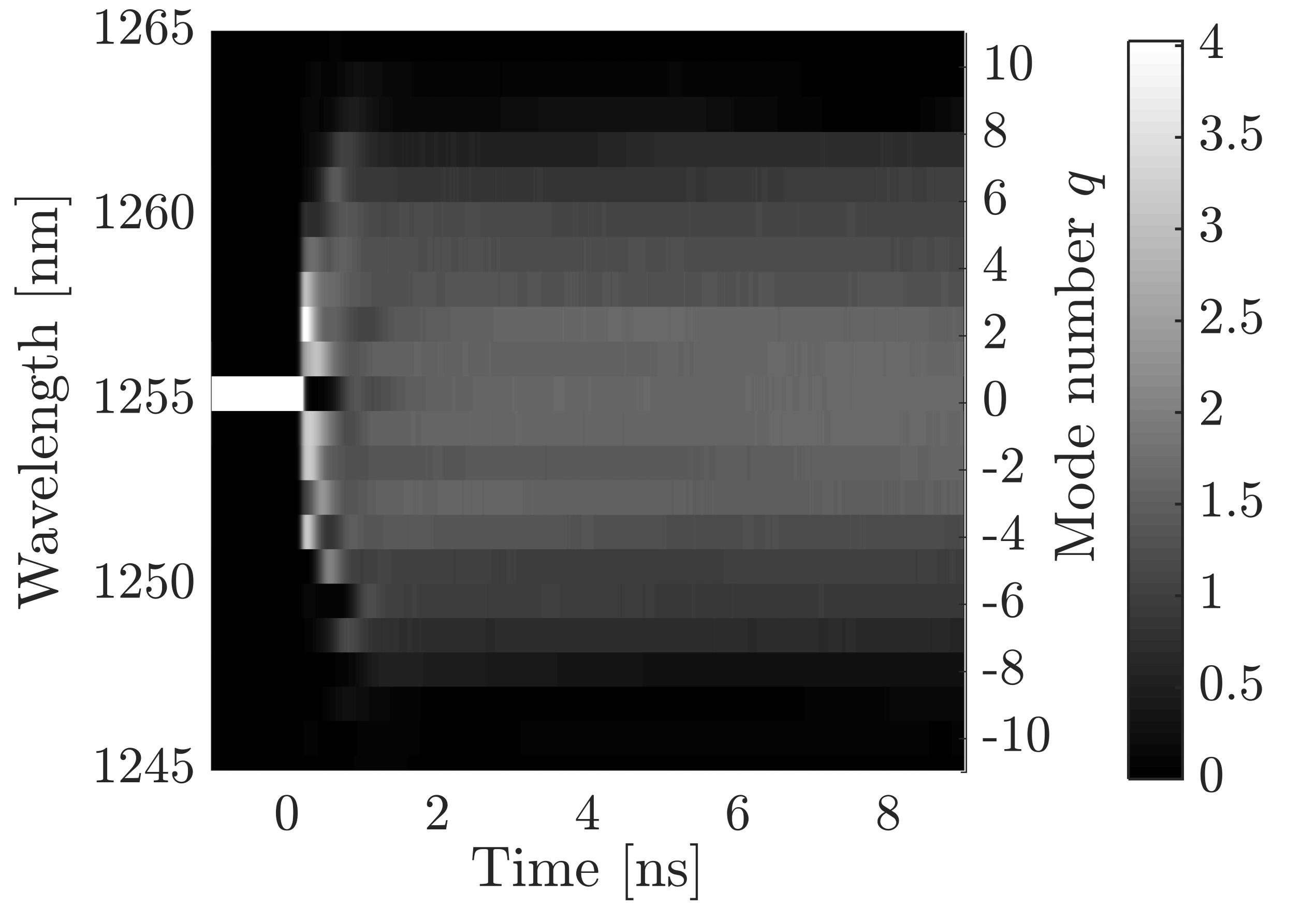}}
	\hspace{0mm}
	\subfigure[]
	{\includegraphics[height=4.7cm]{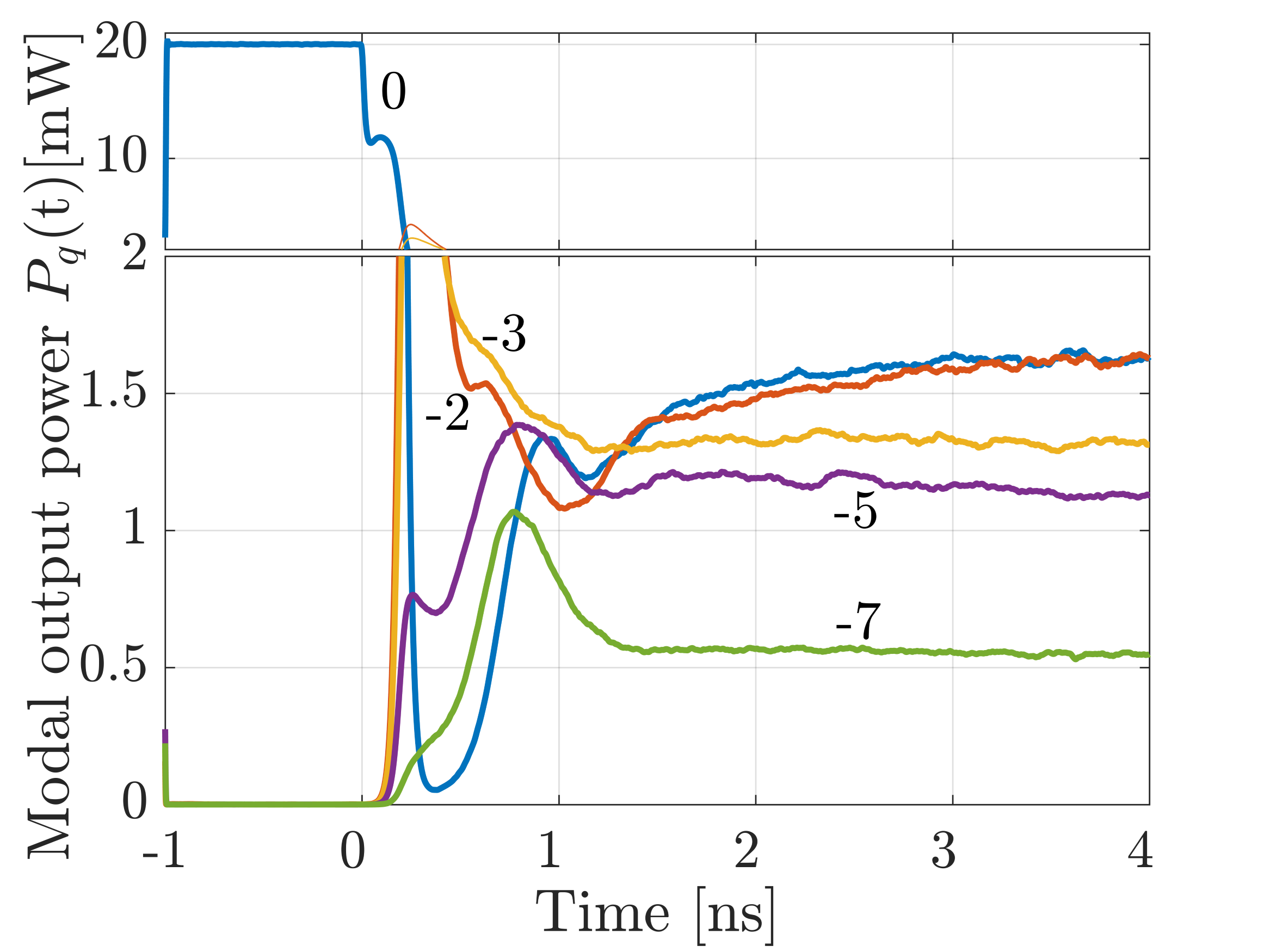}}
	\caption{Temporal evolution of the total optical power ($P(t)$) (a) and of the modal optical power ($P_{q}(t)$) (b,c) obtained by numerical integration of Eqs. (\ref{fieldfasta})-(\ref{pop3fasta}). The simulation shows the effect of SHB, switched on at $t=0$, in the destabilization of the single mode lasing emission ($q=0$) and the excitation of sides modes ($q \neq0$).}
	\label{fig1}
\end{figure}

\begin{figure}[ht!]
	\centering
	\includegraphics[height=5.7cm]{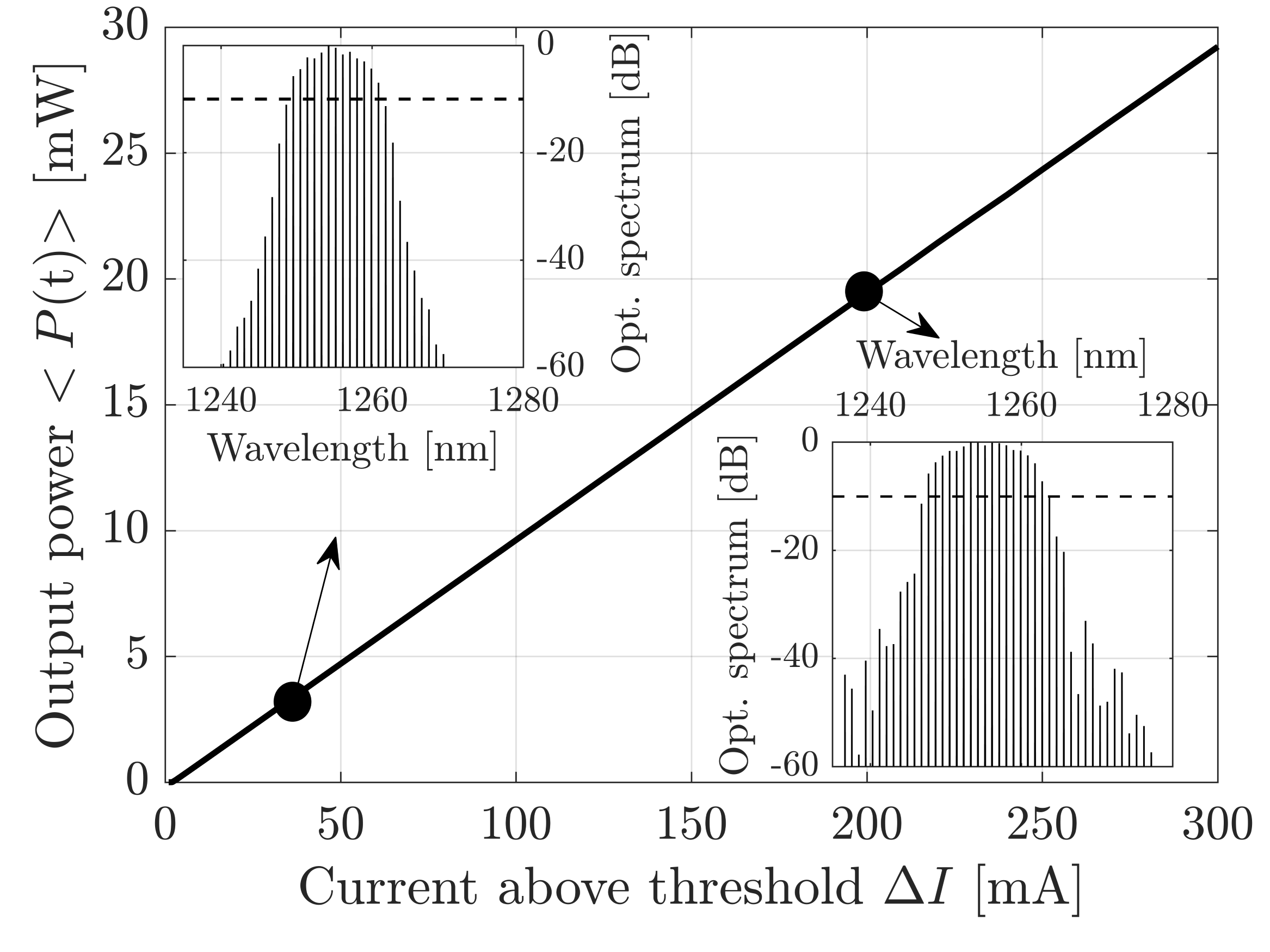}
	\caption{Curve of the average optical power {\it vs.} bias current for the simulated QD laser. The insets show representative optical spectra for $\Delta I$ $=$ \SI{40}{\milli\ampere} and $\Delta I$ $=$\SI{200}{\milli\ampere}.}
	\label{fig2}
\end{figure}

 To study the effect of the carrier grating, in Fig.\,\ref{fig1} we have prepared the laser with an injected current of \SI{200}{\milli\ampere} above the threshold current and, for $t<0$, we have forced $\rho_{m}^{+}=0$ with $m=GS, ES, WL$ thus neglecting Eqs. (\ref{pop2fasta}), (\ref{popESfasta}) and (\ref{pop3fasta}). As expected, the system has a stable Continuous Wave (CW) solution consisting in only one lasing mode ($q=0$) that is the one closest to gain peak \cite{Moloney1, Gioannini}. At $t\geq0$ we ``switch-on'' the fast carrier grating in the system. We plot the temporal evolution of $P(t)$ in Fig.\,\ref{fig1}a,  $P_{q}(t)$ in Fig.\,\ref{fig1}b, and a zoom of the transient of some exemplar optical lines in Fig.\,\ref{fig1}c.
The standing wave pattern induced by the lasing mode $q=0$ generates a carrier grating at the wavelength scale which has two main effects. First, it compresses the gain of the mode $q=0$, triggering a significant reduction of its modal amplitude (as shown in Fig.\,\ref{fig1}b,c) and a consequent increase of the carrier density ($\rho_{0, m}$, with $m = GS, ES, WL$) above threshold (SHB). 
Second, this spatial modulation of the carriers allows the side lines gain ($q \neq 0$) to overcome the cavity losses in a cascading mechanism that brings lines with larger distance from the central one to be later excited (Fig.\,\ref{fig1}b,c).
In agreement with the experimental findings (see e.g. \cite{Lu}), we observe that SHB makes the laser operating  in multi-wavelength regime just above the lasing threshold.  Figure \ref{fig2} shows the power versus current curve for the simulated QD laser from threshold ($\Delta I=I-I_{th}=$ \SI{0}{\milli\ampere}) to $\Delta I=$ \SI{400}{\milli\ampere} and, in the insets, two representative optical spectra corresponding to $\Delta I=$ \SI{40}{\milli\ampere} and $\Delta I=$ \SI{200}{\milli\ampere}. In this current range  we see an increment of the \SI{-10}{\decibel} optical bandwidth from \SI{\sim 14}{\nano\meter} (just above threshold) to \SI{\sim 18}{\nano\meter}.\\


After having identified in the SHB the mechanism responsible for the multi-wavelength lasing, we analyse in the rest of this Section two different dynamical regimes encountered by increasing the current above threshold: a multi-mode regime with phase-unlocked modes (indicated in the following as unlocked regime) and a multi-wavelength regime with phase-locked modes (indicated in the following as locked regime). The latter corresponds to the actual OFC generation. These regimes will be mainly characterized by mapping the temporal evolution of the quantities $P_{q}(t)$, $\Delta \Phi_{q}(t)$ and estimating the OFC indicators defined in the previous paragraph.

\subsection{Unlocked regime}
\begin{figure}[ht!]
	\centering
	\subfigure[]
	{\includegraphics[height=4.7cm]{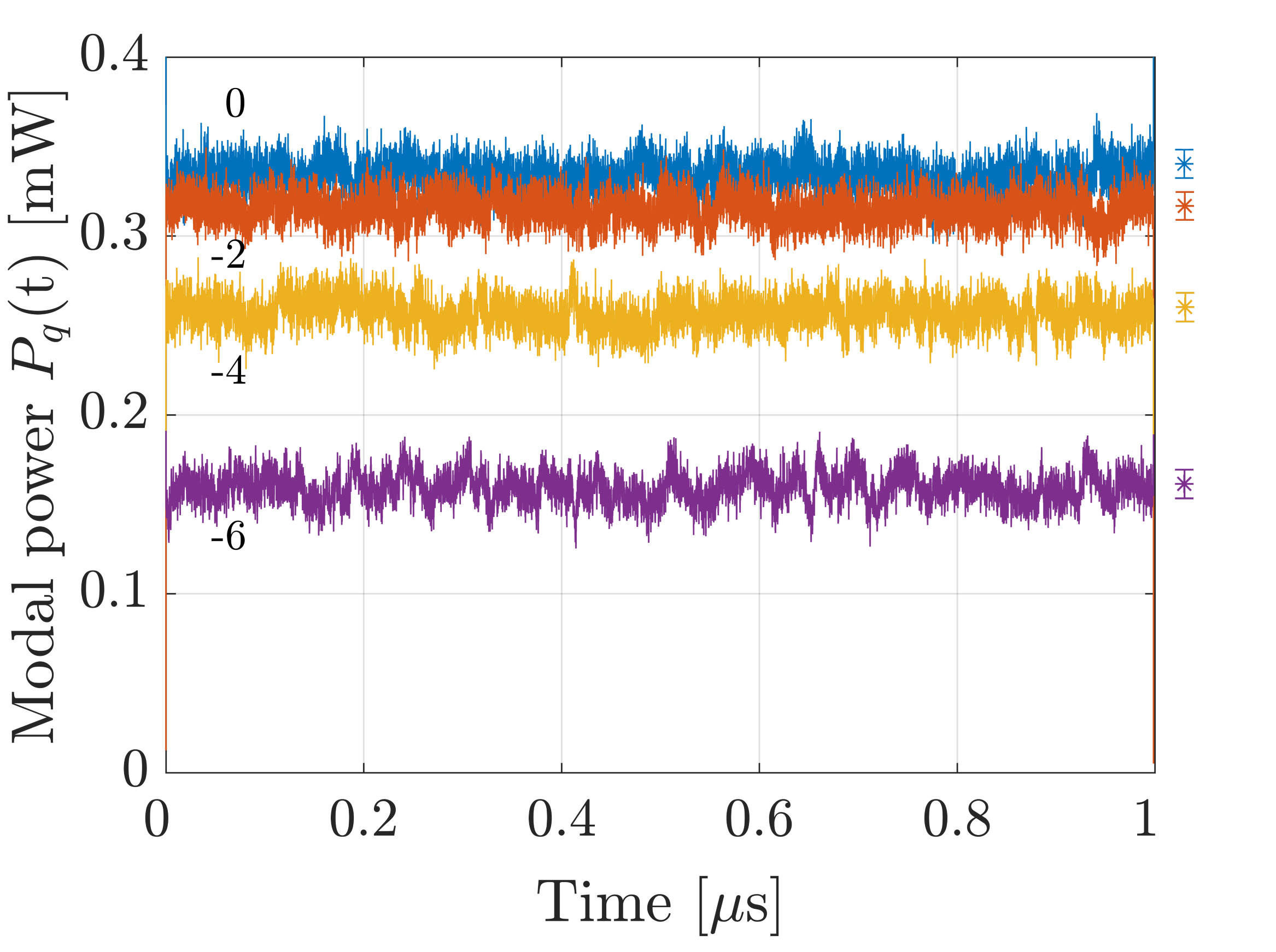}}
	\hspace{0mm}
	\subfigure[]
	{\includegraphics[height=4.7cm]{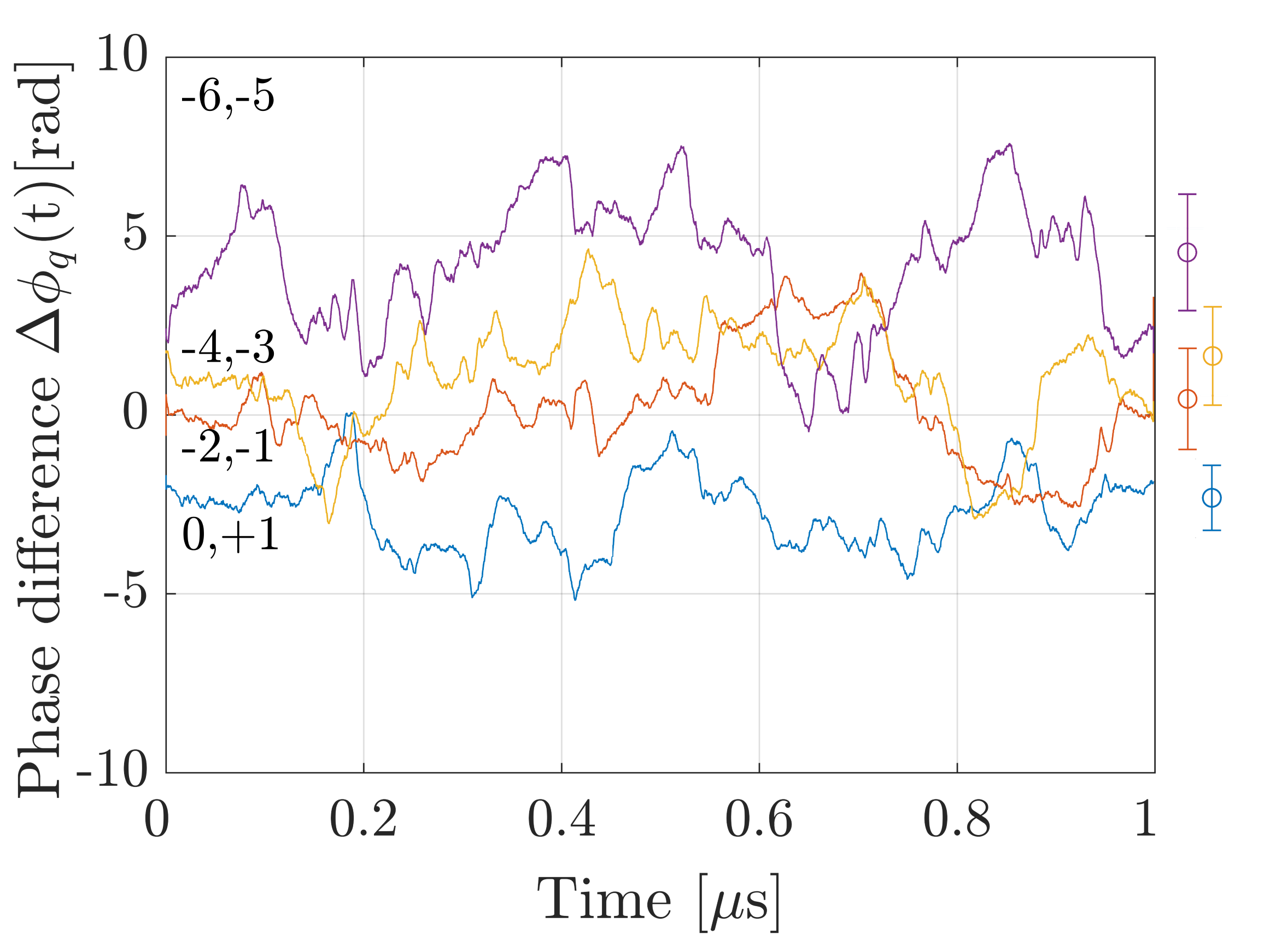}}
	\hspace{0mm}
	\subfigure[]
	{\includegraphics[height=4.7cm]{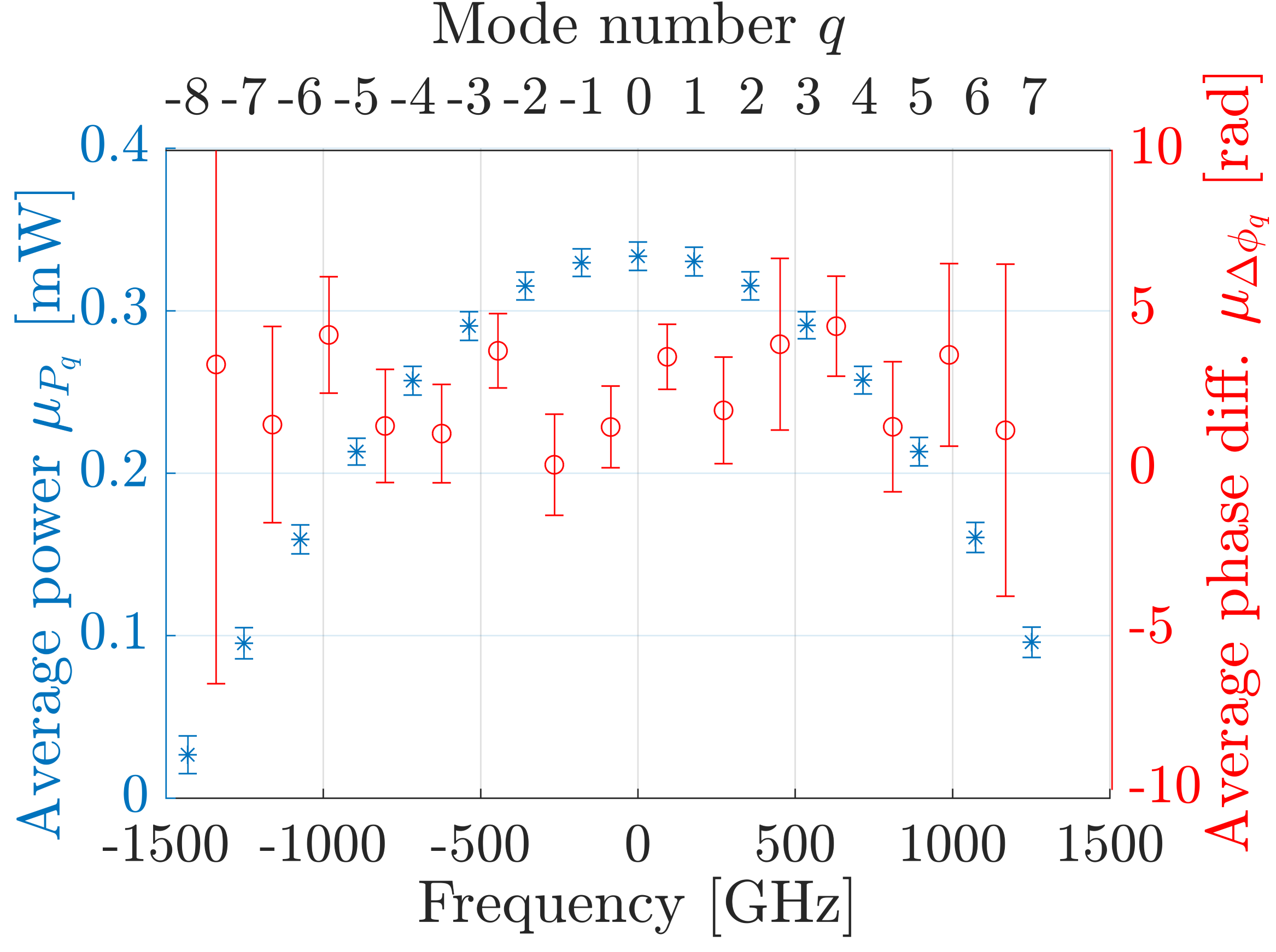}}
	\subfigure[]
	{\includegraphics[height=4.7cm]{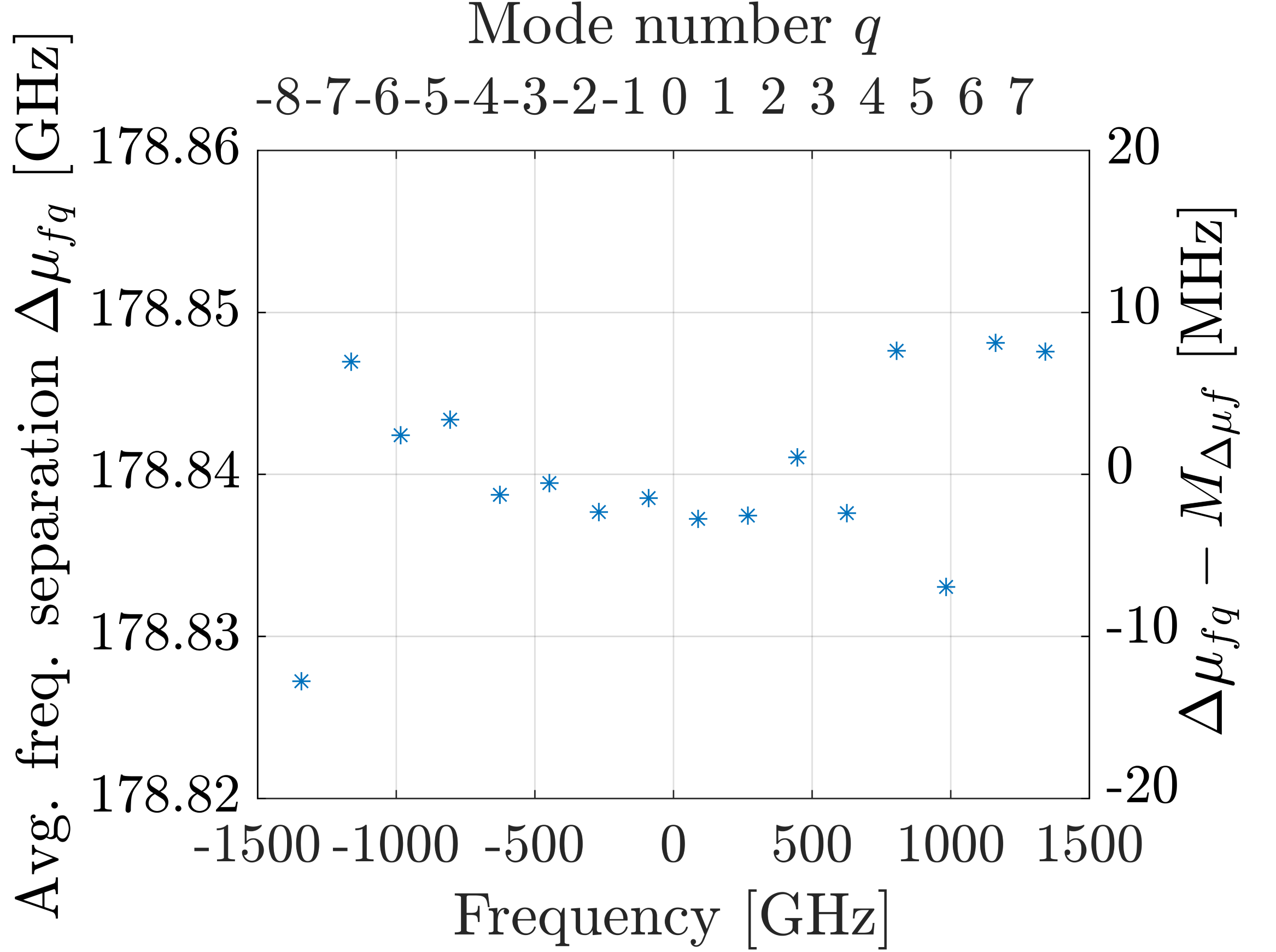}}
	\caption{Unlocked multi-mode dynamics for a pump current \SI{40}{\milli\ampere}. Temporal evolution of the modal power ($P_{q}(t)$) (a) and of phase difference between adjacent modes ($\Delta \Phi_{q}(t)$) (b) for selected modes. The average values ($\mu_{Pq}$ and $\mu_{\Delta \Phi q}$) and standard deviations ($\sigma_{\delta P_q}$ and $\sigma_{\Delta \Phi q}$) are reported next to the right vertical axis. In panel (c) these statistical moments are reported for all the modes within \SI{-10}{\decibel} bandwidth in the optical spectrum. The temporal average of the frequency separations between couples of adjacent modes ($\Delta \mu_{f q}$) (left vertical axis) and its difference with respect to the overall average over the selected modes ($M_{\Delta \mu_{f}}$) (right vertical axis) are represented in panel (d).}
	\label{fig3}
\end{figure}

As an example, we report the results of an unlocked regime observed at \SI{40}{\milli\ampere} above threshold.  Figure \ref{fig3} shows $P_{q}(t)$ and  $\Delta \Phi_{q}(t)$ for a few representative lines of the optical spectrum; we plot next to the right vertical axis also the temporal averages $\mu_{Pq}$ and $\mu_{\Delta \Phi q}$ (circular symbols) and the associated intensity noise $\sigma_{\delta P_q}$ and differential phase noise $\sigma_{\Delta \Phi q}$  (bars amplitudes). All the parameters have been calculated for an observation time window of \SI{1}{\micro\second}.
The same statistical moments are also plotted in Fig.\,\ref{fig3}c for the lasing lines within the \SI{-10}{\decibel} optical bandwidth and, together with the average frequency separation between couples of adjacent lines ($\Delta \mu_{f q}$ plotted in Fig.\,\ref{fig3}d), they represent convenient indicators of the self-generated OFCs.
The evidence that differential phase fluctuations are uncorrelated in time, that
differential phase standard deviation bars are quite large ($> 1$ rad/s) for all lines and that optical frequency separation is quite dependent on the mode number allows us to conclude that we are dealing with an unlocked regime.

\subsection{Locked regime: self-generated OFC}

\begin{figure}[ht!]
	\centering
	\subfigure[]
	{\includegraphics[height=4.7cm]{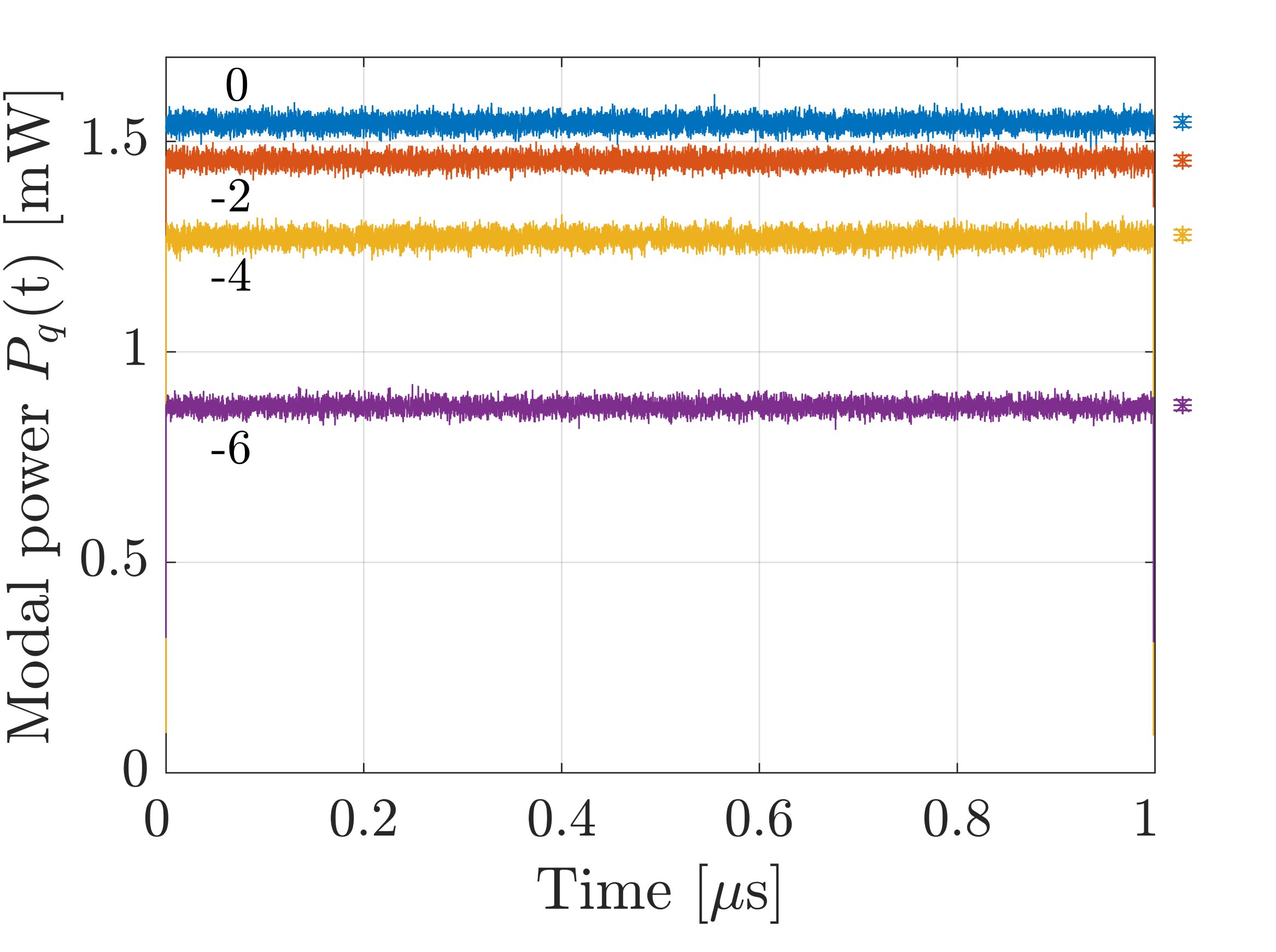}}
	\hspace{0mm}
	\subfigure[]
	{\includegraphics[height=4.7cm]{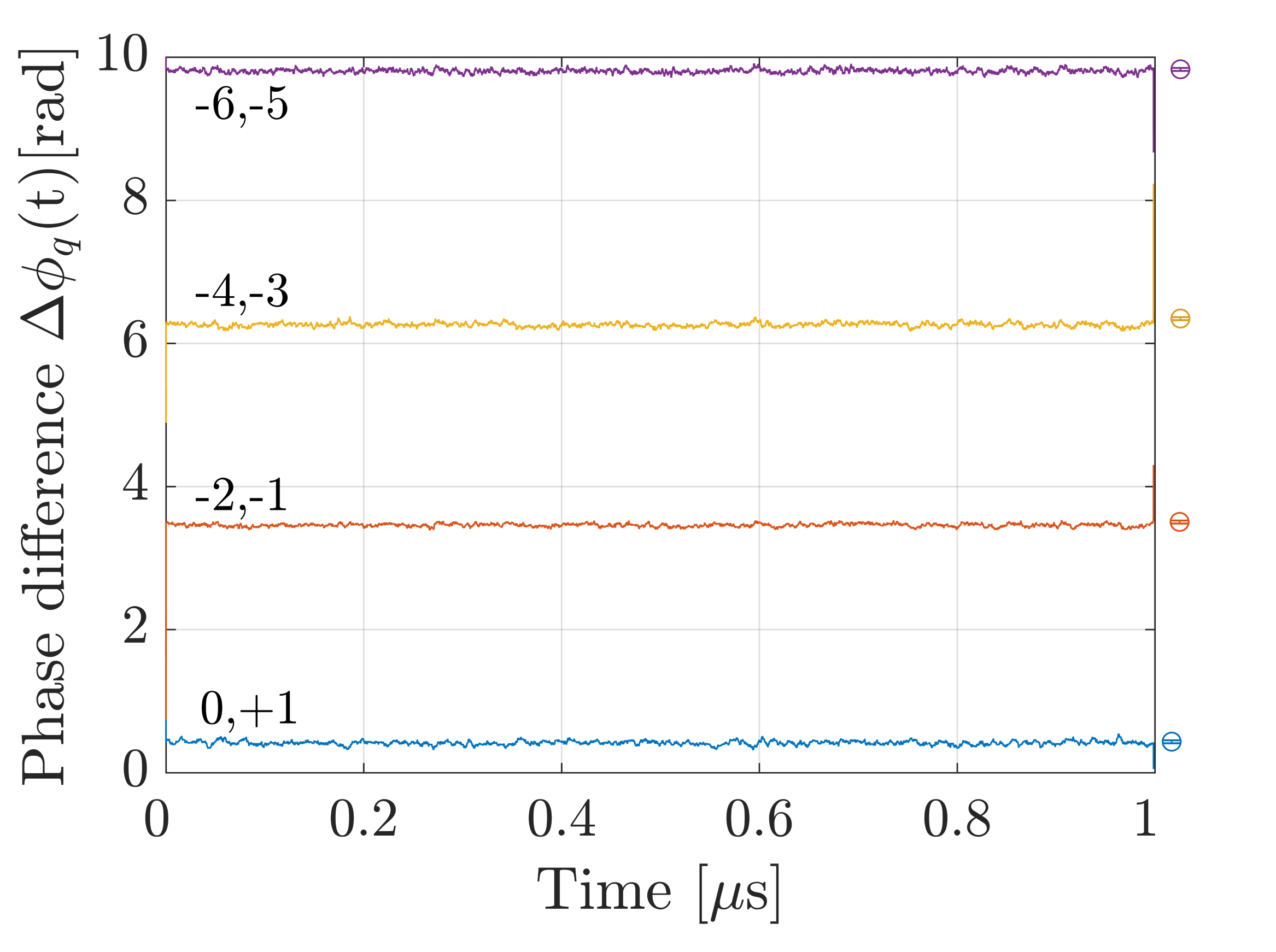}}
	\hspace{0mm}
	\subfigure[]
	{\includegraphics[height=4.7cm]{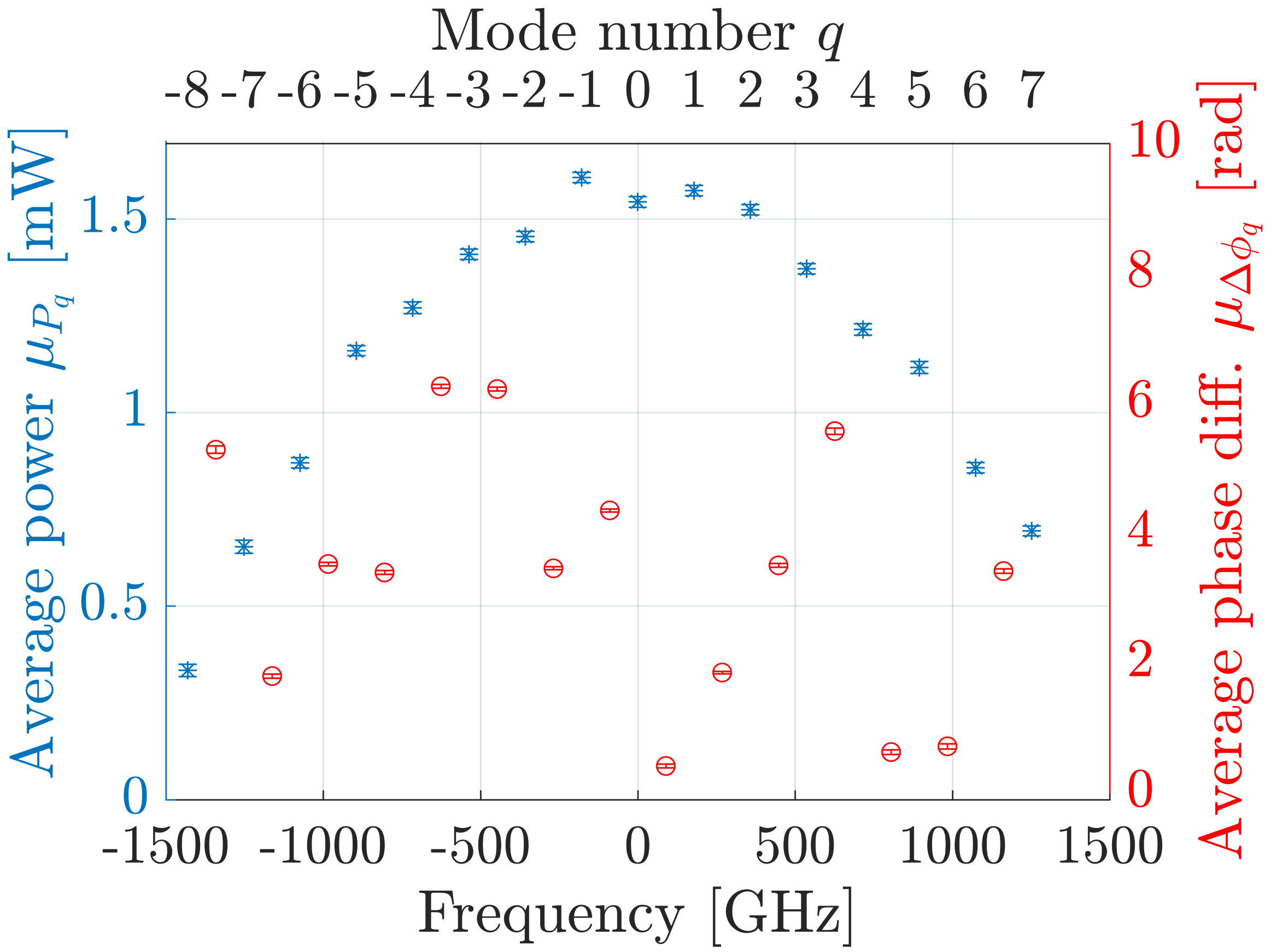}}
	\subfigure[]
	{\includegraphics[height=4.7cm]{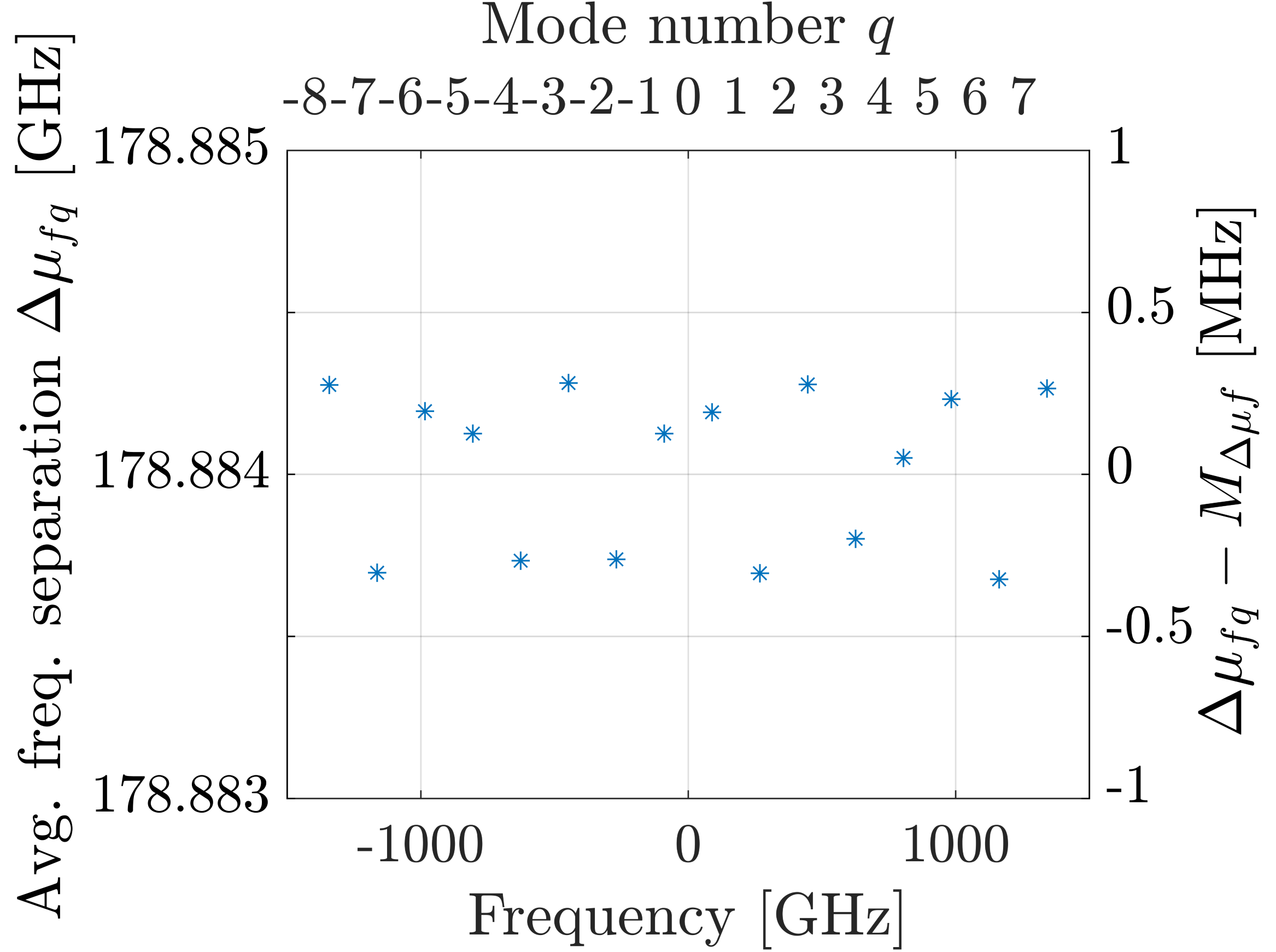}}
	\caption{Self-generated OFC regime for a pump current of \SI{200}{\milli\ampere}. Temporal evolution of the modal power  ($P_{q}(t)$) (a) and of phase difference between adjacent modes ($\Delta \Phi_{q}(t)$) (b) for selected modes. The average values ($\mu_{Pq}$ and $\mu_{\Delta \Phi q}$) and standard deviations ($\sigma_{\delta P_q}$ and $\sigma_{\Delta \Phi q}$) are reported next to the right vertical axis. In panel (c) these statistical moments are reported for all the modes within \SI{-10}{\decibel} bandwidth in the optical spectrum. The temporal average of the frequency separations between couples of adjacent modes ($\Delta \mu_{f q}$) (left vertical axis) and its difference with respect to the overall average over the selected modes ($M_{\Delta \mu_{f}}$) (right vertical axis) are represented in panel (d).}
	\label{fig4}
\end{figure}

For higher bias current, for example \SI{200}{\milli\ampere} above threshold, the system dynamics changes radically as shown in Fig.\,\ref{fig4}. Compared to the case in Fig.\,\ref{fig3}, we see a consistent reduction of both the intensity noise and the differential phase fluctuations and a significant uniformity of the optical line spacing; all indicating that the phase-locking with OFC generation has been achieved. Since the differential phases are practically constant with time, but not the same for all the lines, it is not possible to observe any optical pulse at the laser output (as it happens in passive mode-locking) despite that the modes are phase-locked.
\\

\begin{figure}[ht!]
	\centering
	\subfigure[]
	{\includegraphics[height=4.7cm]{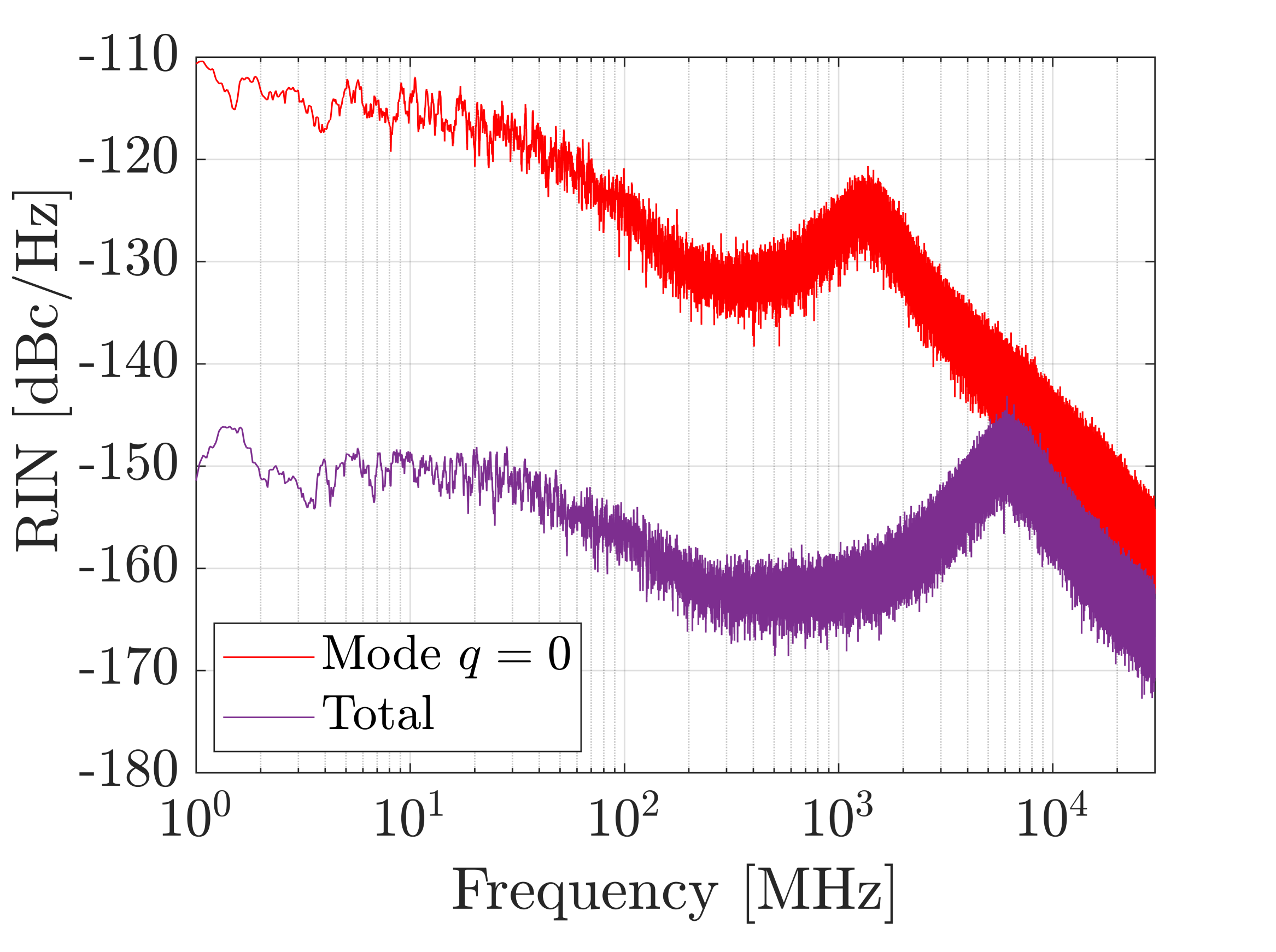}}
	\hspace{0mm}
	\subfigure[]
	{\includegraphics[height=4.7cm]{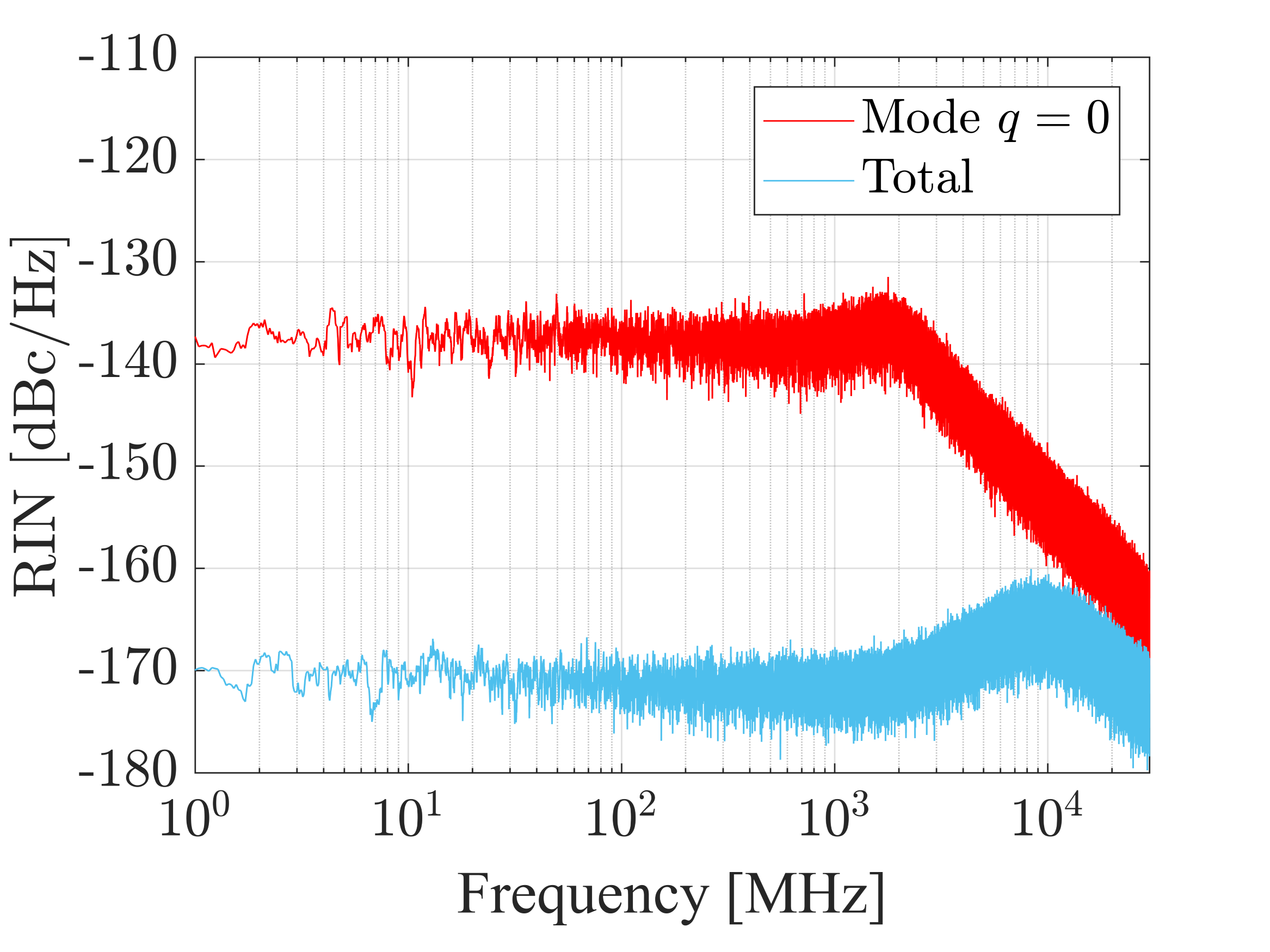}}
	\hspace{0mm}
		\subfigure[]
	{\includegraphics[height=4.7cm]{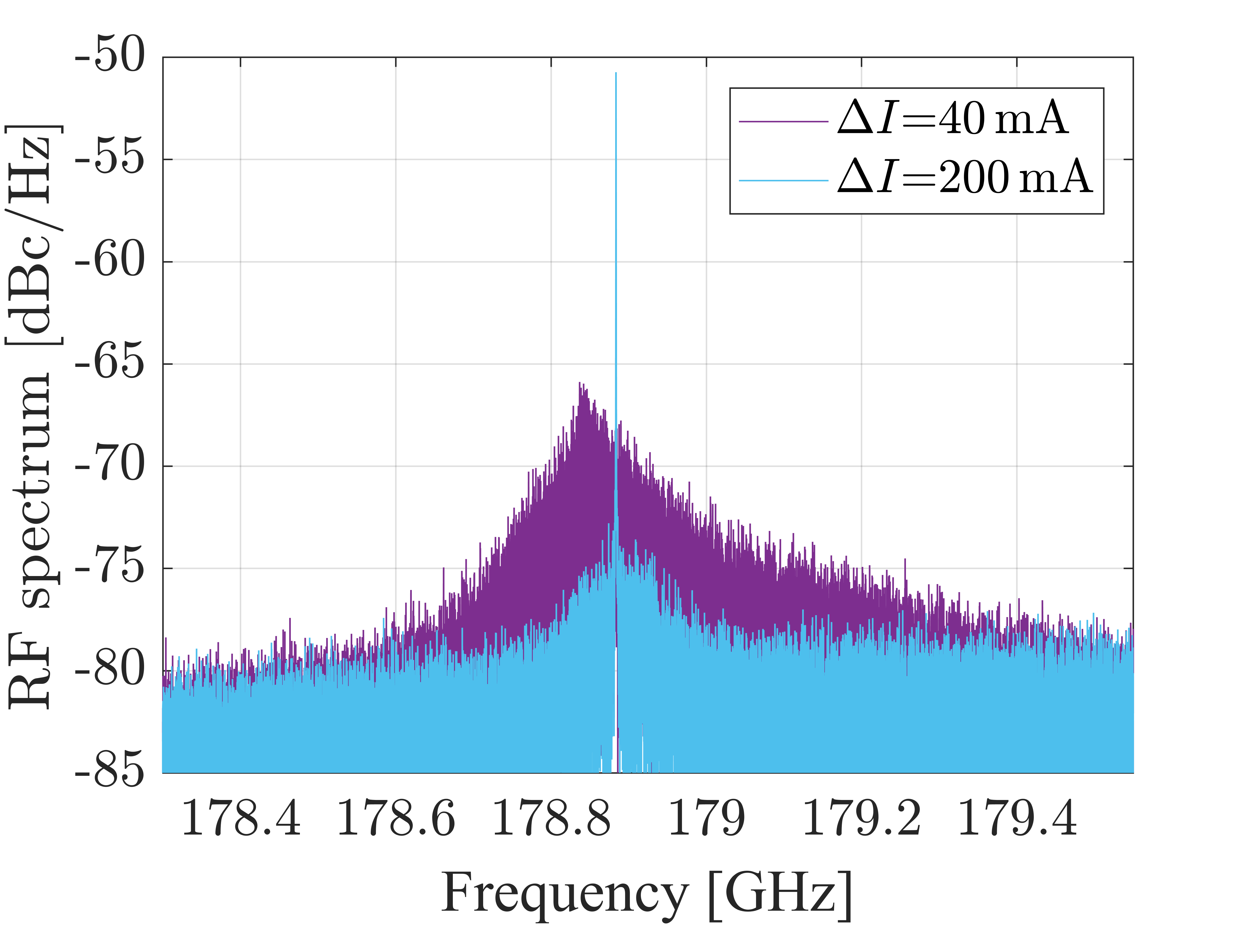}}
	\hspace{0mm}
	\subfigure[]
	{\includegraphics[height=4.7cm]{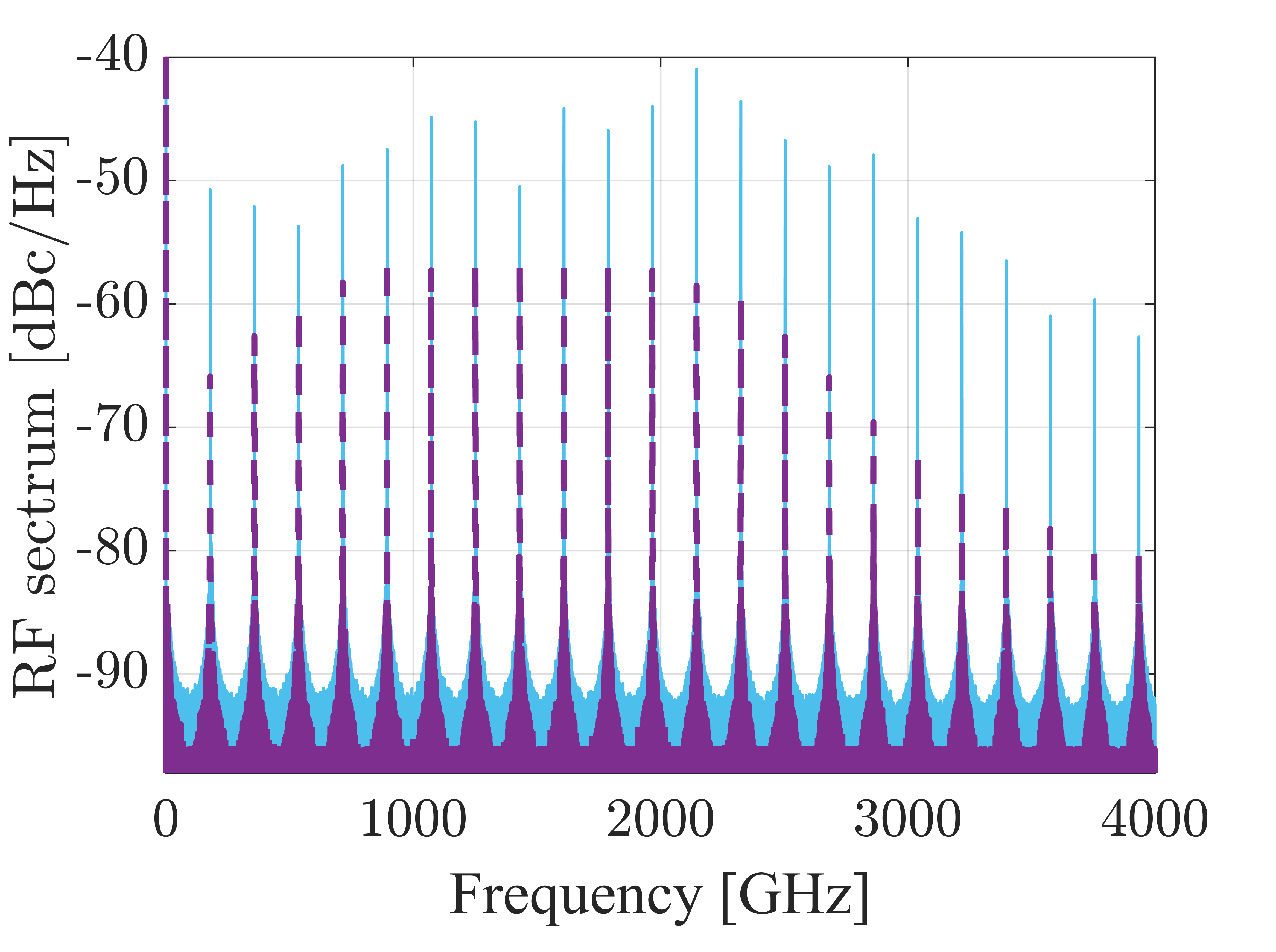}}
	\caption{RIN spectra associated to the total power (violet/cyan lines) and the modal power of the central mode in the optical spectrum ($q=0$) (red lines) for $\Delta I=$\SI{40}{\milli\ampere} (a) and $\Delta =$ \SI{200}{\milli\ampere} (b). First beat-note (c) and total power spectrum (d) for $\Delta I=$ \SI{40}{\milli\ampere} (violet line) and $\Delta I=$ \SI{200}{\milli\ampere} (cyan line).}
	\label{fig3bis}
\end{figure}

We compare the two regimes (locked and unlocked at \SI{200}{\milli\ampere} and \SI{40}{\milli\ampere} above threshold respectively) in terms of degree of coherence of the system by reporting quantifiers which can be more easily measured: the RIN spectra (Fig.\,\ref{fig3bis}a,b), the beat-note RF linewidth (Fig.\,\ref{fig3bis}c), and the RF spectrum in an extended frequency range (Fig.\,\ref{fig3bis}d). In the unlocked regime we observe high RIN (especially in the low frequency part of the spectrum between \SI{1}{\mega\hertz} and \SI{1}{\giga\hertz}) and very broad RF line at the beat-note. On the contrary, the locked regime is characterized by a reduced RIN (it decrease down to around \SI{-170}{\decibelc\per\hertz} for total power RIN and to around \SI{-140}{\decibelc\per\hertz} for RIN of mode $q=0$) and extremely narrow beat-note. Finally, the RF spectrum in a broader frequency range reveals in both cases the presence of more than $10$ lines in the first decade.\\

The temporal evolution of the total power in a locked regime is dominated by regular oscillations with \SI{\sim 5}{\pico\second} period (corresponding to the inverse of the beat-note) superimposed to sub picosecond oscillations associated with the high frequency peak around  \SI{2}{\tera\hertz} in the RF spectrum of Fig.\,\ref{fig3bis}d. We associate this high frequency peak with the Rabi resonance \cite{Lugiato} and we consider it as a signature of the coherent interaction between the intracavity electric field and the ensemble of ``artificial atoms'' represented by the QDs in the active material . Experimental and theoretical evidences of Rabi oscillations at room temperature in QD materials have been found in QD based semiconductor optical amplifier in condition of high intensity pulse injection \cite{Kolarczik,Capua} but, at the best of our knowledge, no signature of this phenomenon was reported in QD laser diodes. We however stress here that the observation of the Rabi resonance is in this context only the consequence of the excitation of the QD medium with the  coherent broad band electric field self-generated by the FWM process, as discussed in the next Section. Therefore, contrary to the prediction of the Risken-Nummedal analysis for a two level system laser \cite{Risken, Lugiato}, the excitation of the Rabi resonance is not the key physical mechanism that triggers the phase-locking of the optical lines. Indeed, if the Risken-Nummendal instability was responsible for the phase-locking, we would have found only locked lasing lines separated by \SI{\sim 2} {\tera\hertz} \cite{ColumboNusod}.\\

\begin{figure}[ht!]
	\centering
	\subfigure[]
	{\includegraphics[height=4.7cm]{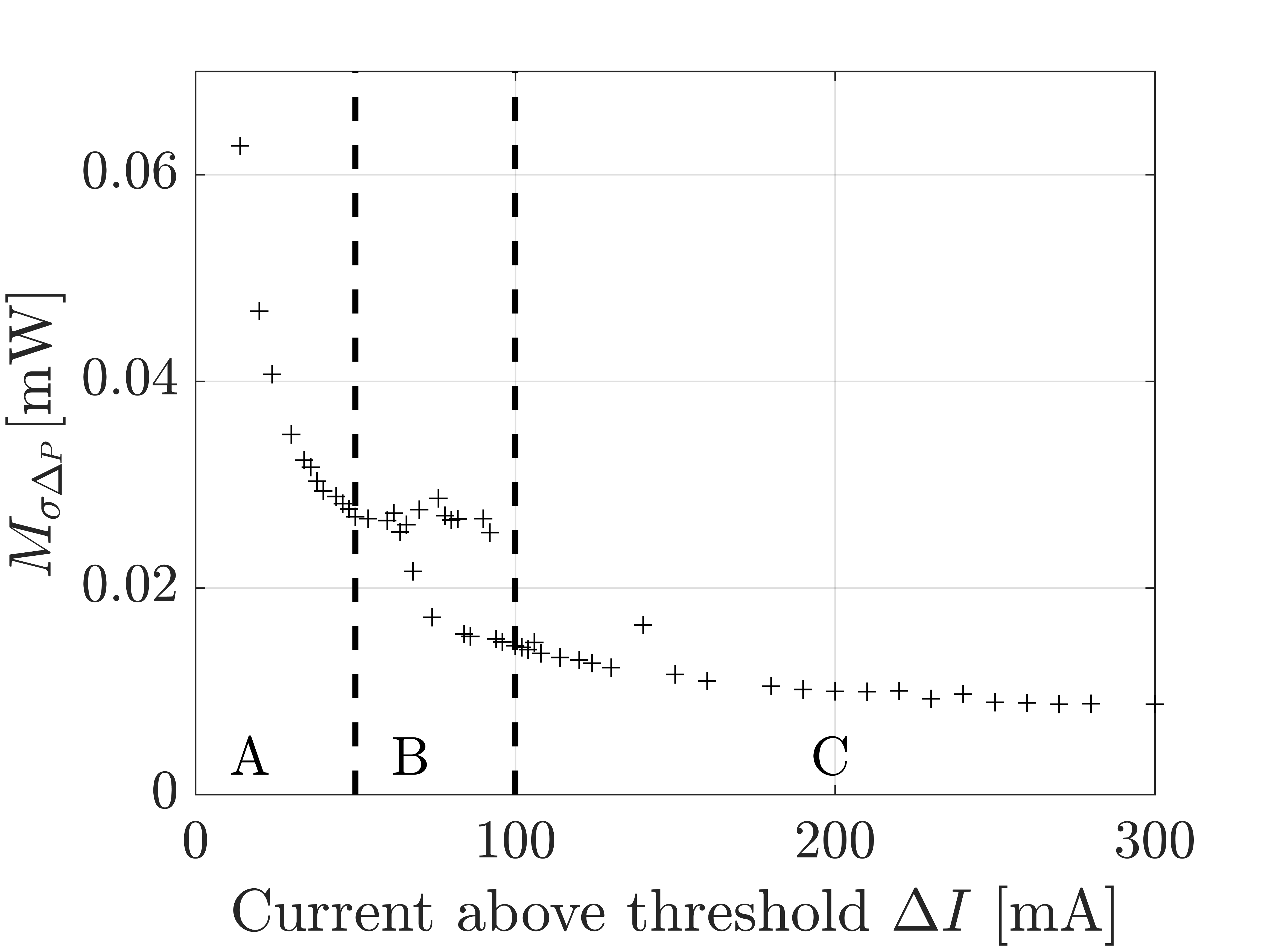}}
	\hspace{0mm}
	\subfigure[]
	{\includegraphics[height=4.7cm]{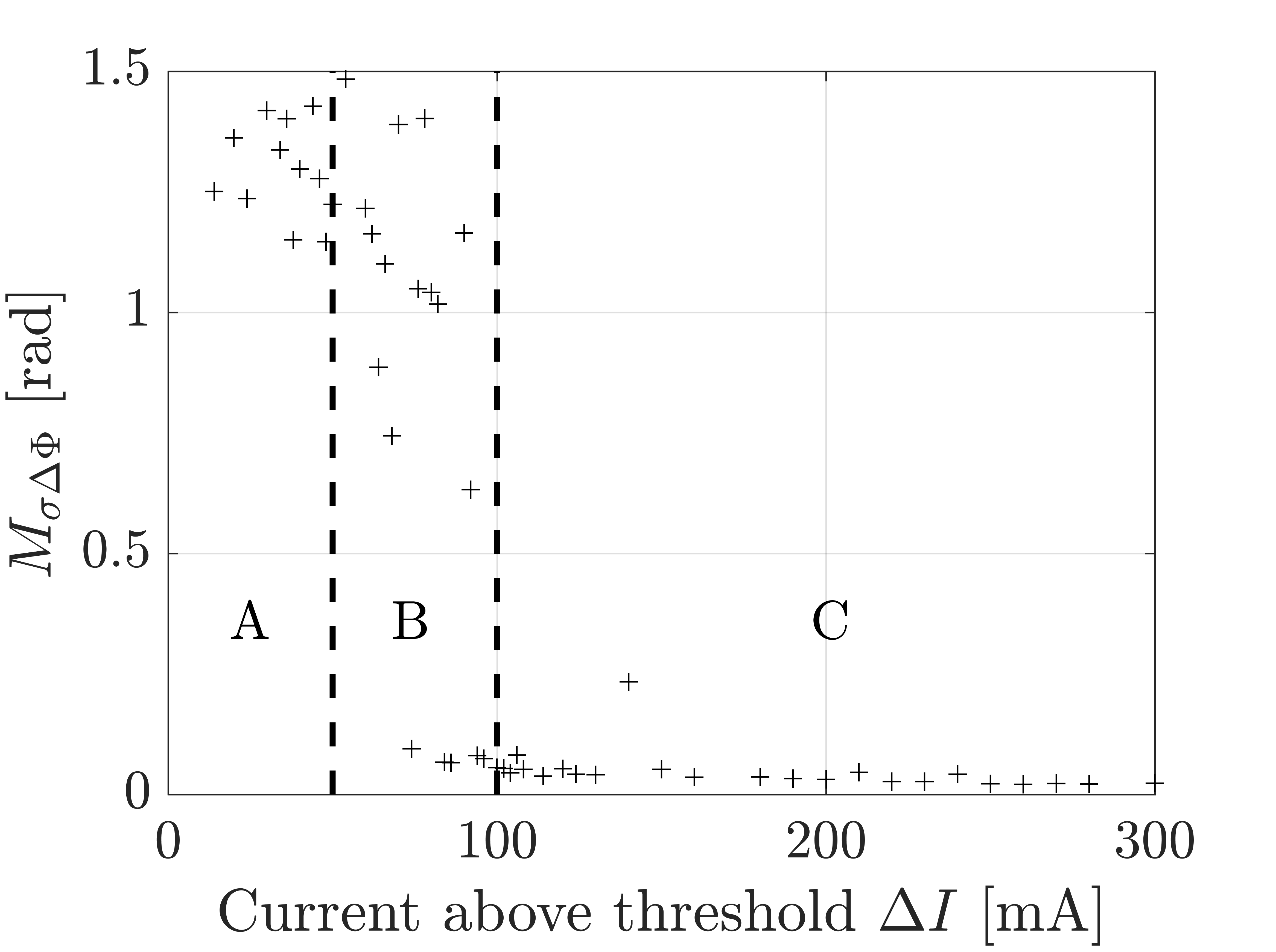}}
	\hspace{0mm}
		\subfigure[]
		{\includegraphics[height=4.7cm]{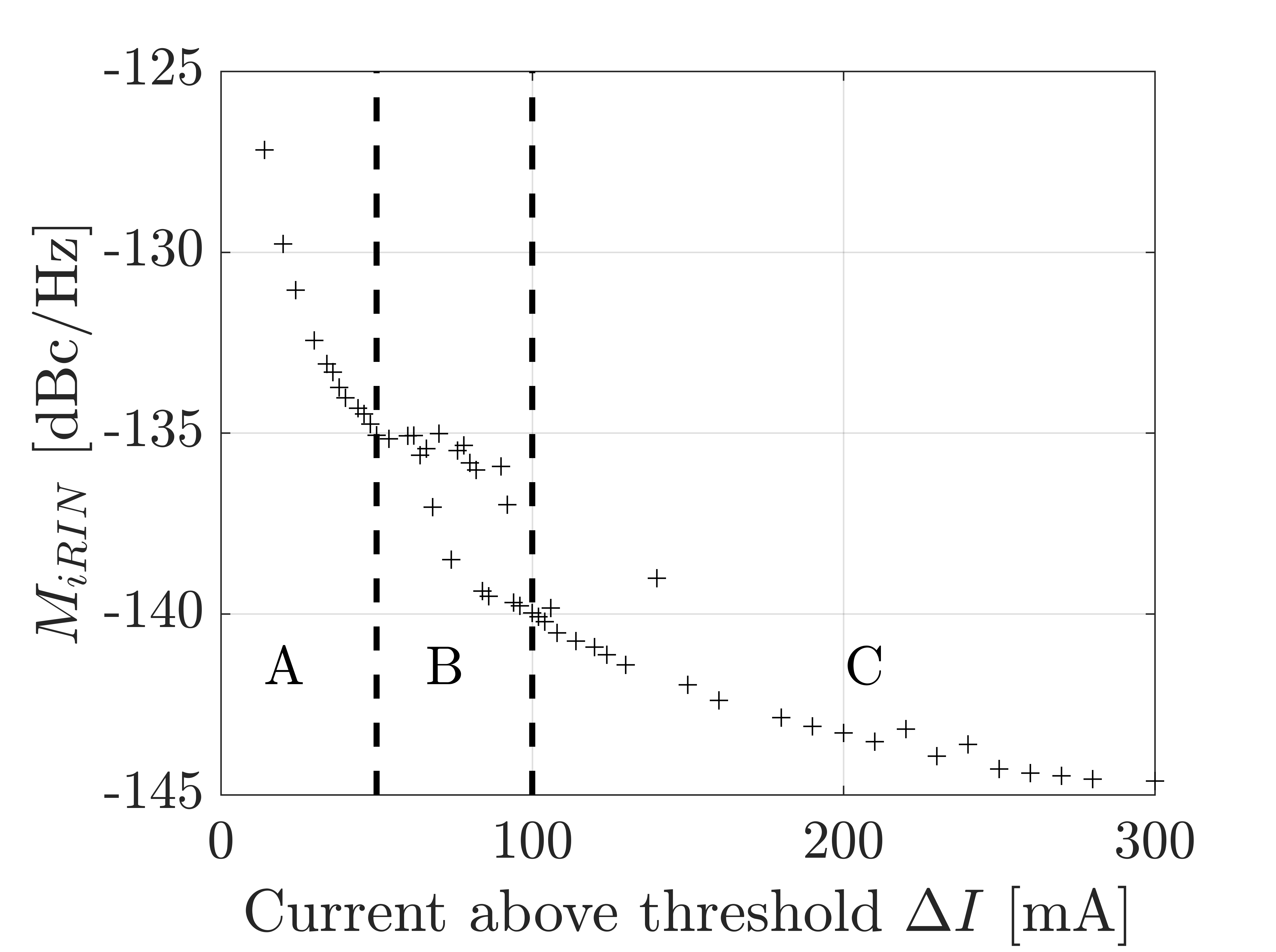}}
	\hspace{0mm}
	{\includegraphics[height=4.7cm]{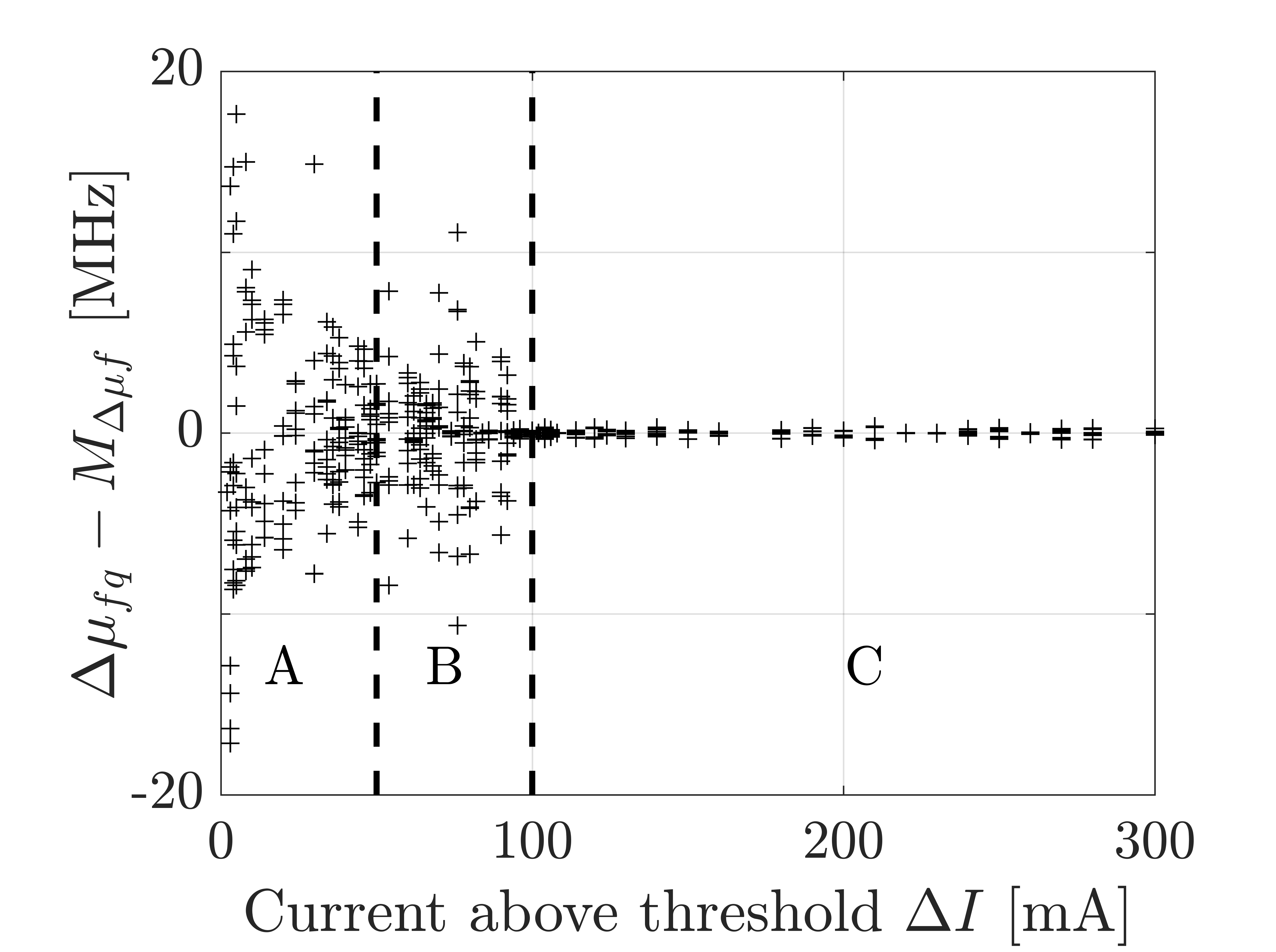}}
	\caption{Average intensity noise ($M_{\sigma_ {\delta P}}$) (a) and average differential phase noise ($M_{\sigma \Delta \Phi}$) (b) for increasing values of the bias current. For the same values we plot the integrated RIN ($M_{iRIN}$) (c) and the average frequency separation between couples of adjacent mode minus its mean over the selected modes ($\Delta\mu_{fq} - M_{\Delta\mu f}$) (d).}
	\label{indi}
\end{figure}
\begin{figure}[ht!]
	\centering
	\includegraphics[height=4.7cm]{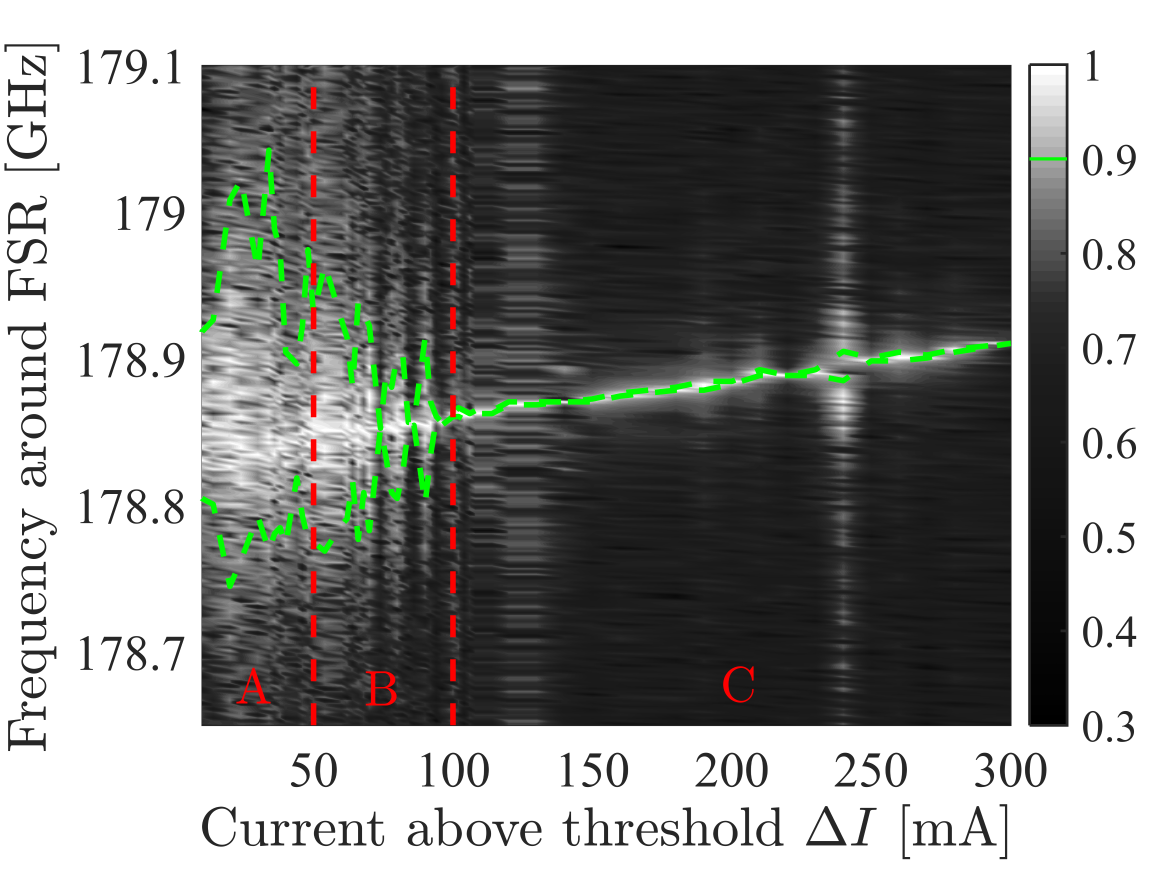}
\caption{Beat note spectrum (normalized for each current, with respect to its maximum) {\it vs.} bias current above threshold. The green dashed line traces the 0.9 contour level} 
\label{indi1}
\end{figure}

In order to describe the transition from unlocked to locked regime, we plot in Fig.\,\ref{indi} the average OFC indicators ($M_{\sigma_ {\delta P}}$, $M_{\sigma\Delta\Phi}$, ...) versus the bias current. The beat-note spectra for the same values of the electrical pumping are reported in the image  of Fig.\,\ref{indi1}.
The typical scenario observed by increasing $\Delta I$ from \SI{0}{\milli\ampere} (laser threshold) to \SI{300}{\milli\ampere} consists initially in a multi-wavelength unlocked regime (denoted with letter $A$ in the figure) associated to large intensity noise (Fig.\,\ref{indi}a and c), large differential phase fluctuations (Fig.\,\ref{indi}b) and very unevenly spaced optical lines (Fig.\,\ref{indi}d). The RF beat-note is also rather  broad (Fig.\,\ref{indi1}). This regime is then followed by clear phase-locking regime (denoted by the letter $C$ in the figure) where both intensity noise and differential phase noise is drastically reduced. We also observe lower integrated RIN, constant optical line spacing and narrow RF beat-note linewidth. 
In the region between $\Delta I$ $=$ \SI{\sim 60}{\milli\ampere} and $\Delta I$ $=$ \SI{\sim 100}{\milli\ampere} (denoted by the letter $B$ in the figure) the two configurations coexist for the same current parameter and the final steady state depends from the system history (initial conditions). After $\Delta I$ $=$ \SI{100}{\milli\ampere} we observe self-generated OFC apart from few values of current where the system is unlocked.
These results well agree with the experimental evidences reported in \cite{Muller} for a single Section QDash laser where it is shown that an increase of the pump current above a certain value brings the laser to move from an unlocked regime (measured as high integrated RIN and large RF line at the beat-note) to a locked regime characterized by a significant drop of the integrated RIN and reduction of the RF linewidth.\\ 

%

\begin{figure}[ht!]
	\centering
	\subfigure[]
	{\includegraphics[height=4.7cm]{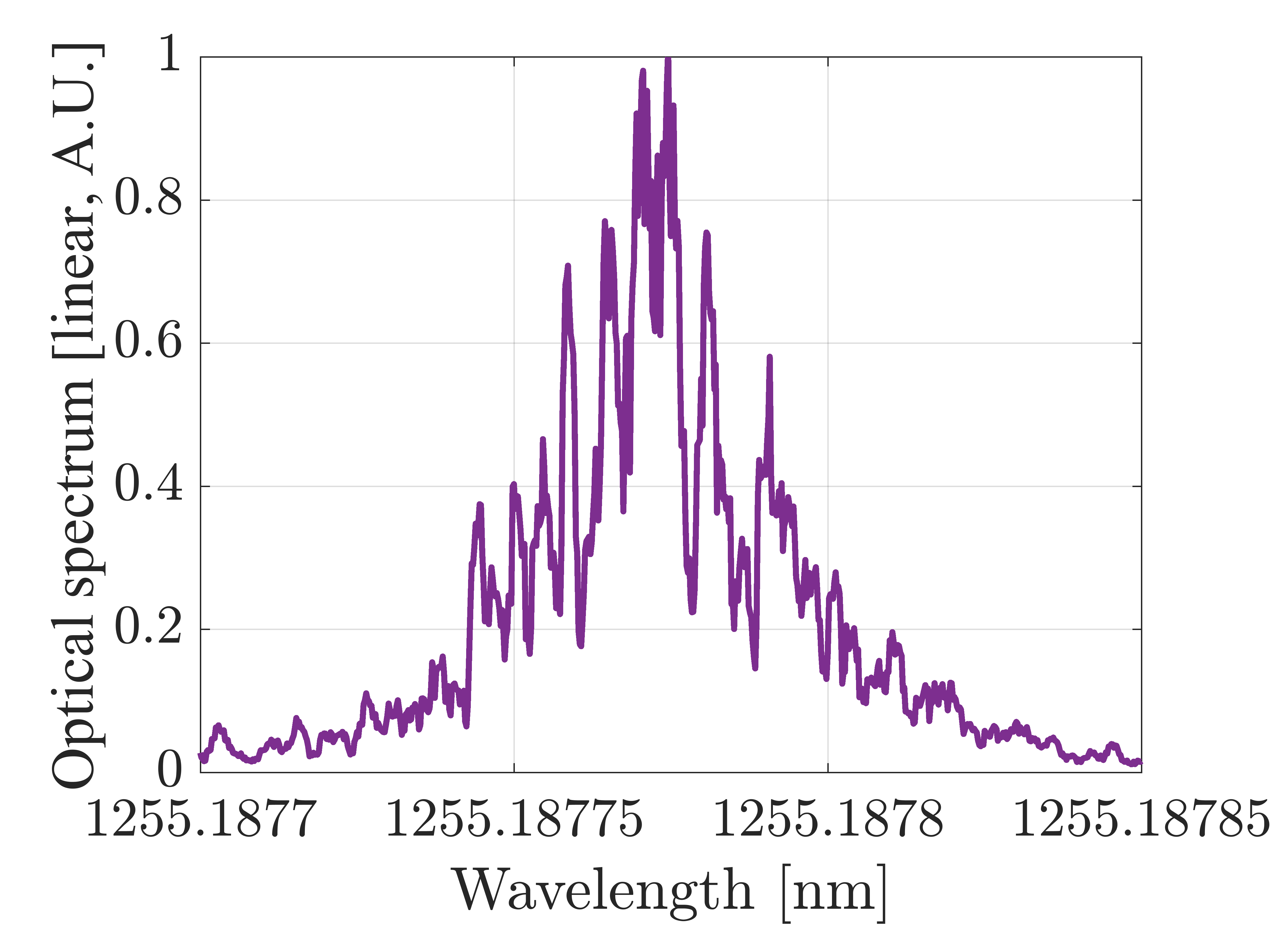}}
	\hspace{0mm}
	\subfigure[]
	{\includegraphics[height=4.7cm]{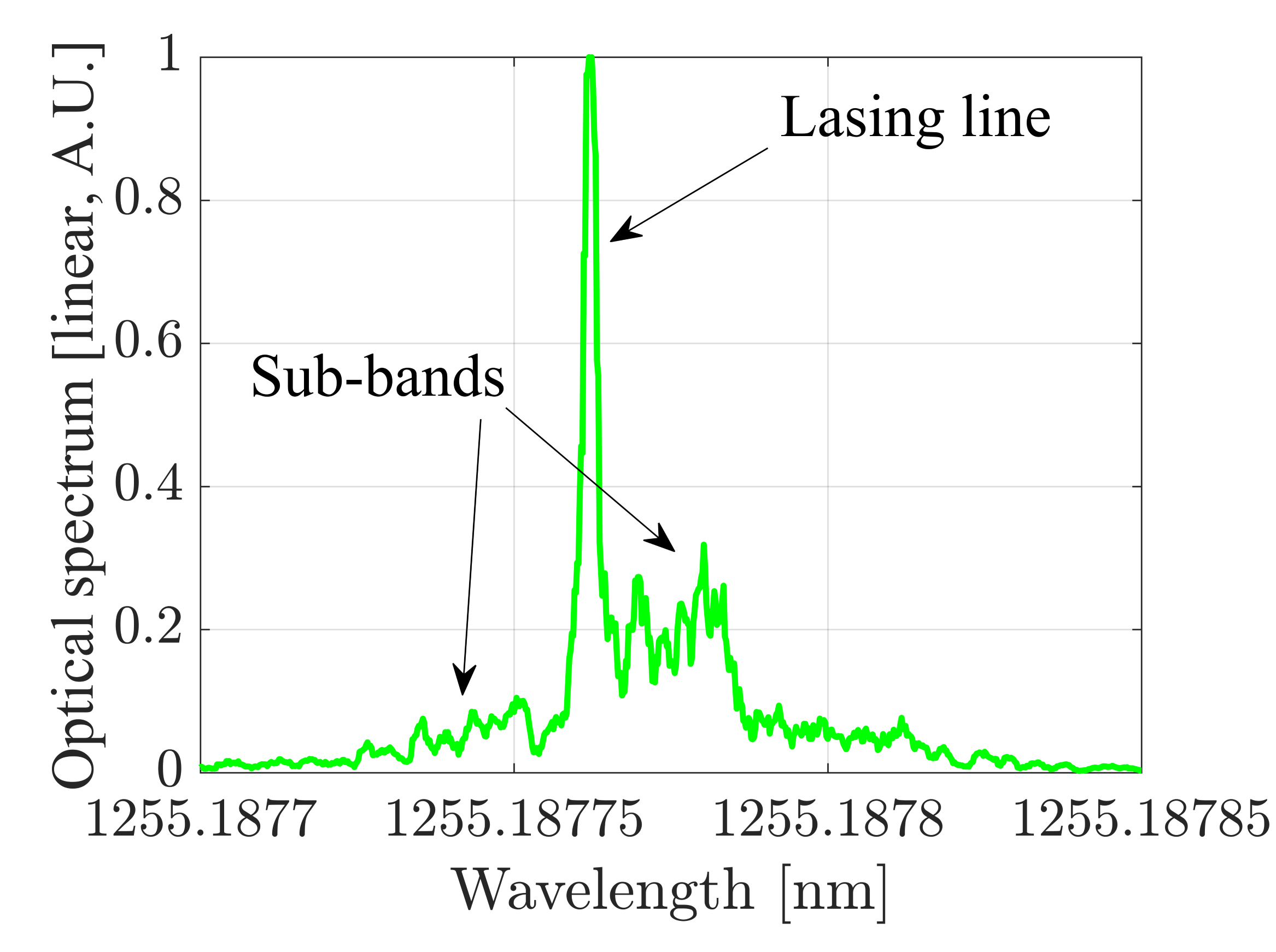}}
	\hspace{0mm}
	\subfigure[]
	{\includegraphics[height=4.7cm]{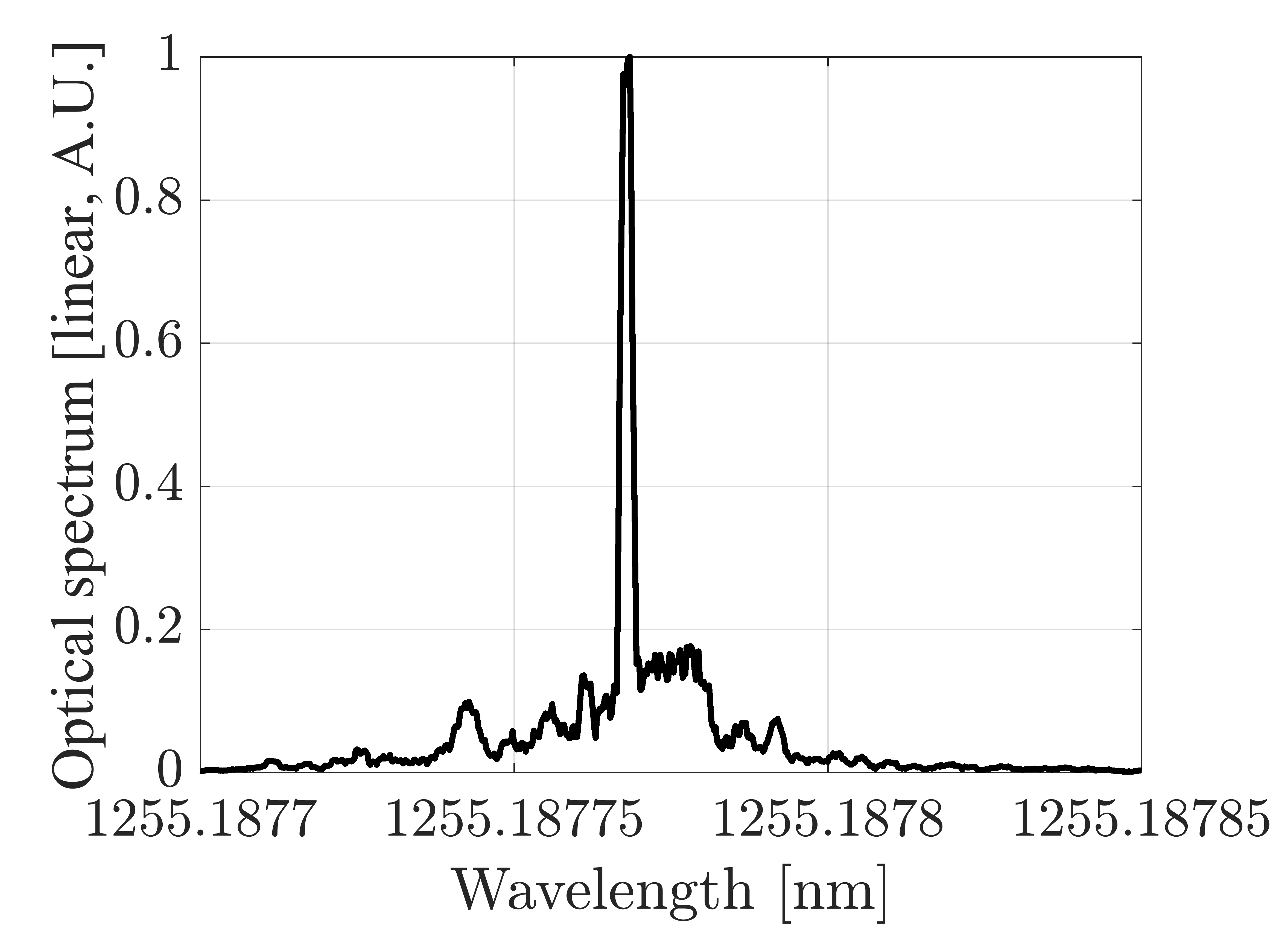}}
	\hspace{0mm}
	\subfigure[]
	{\includegraphics[height=4.7cm]{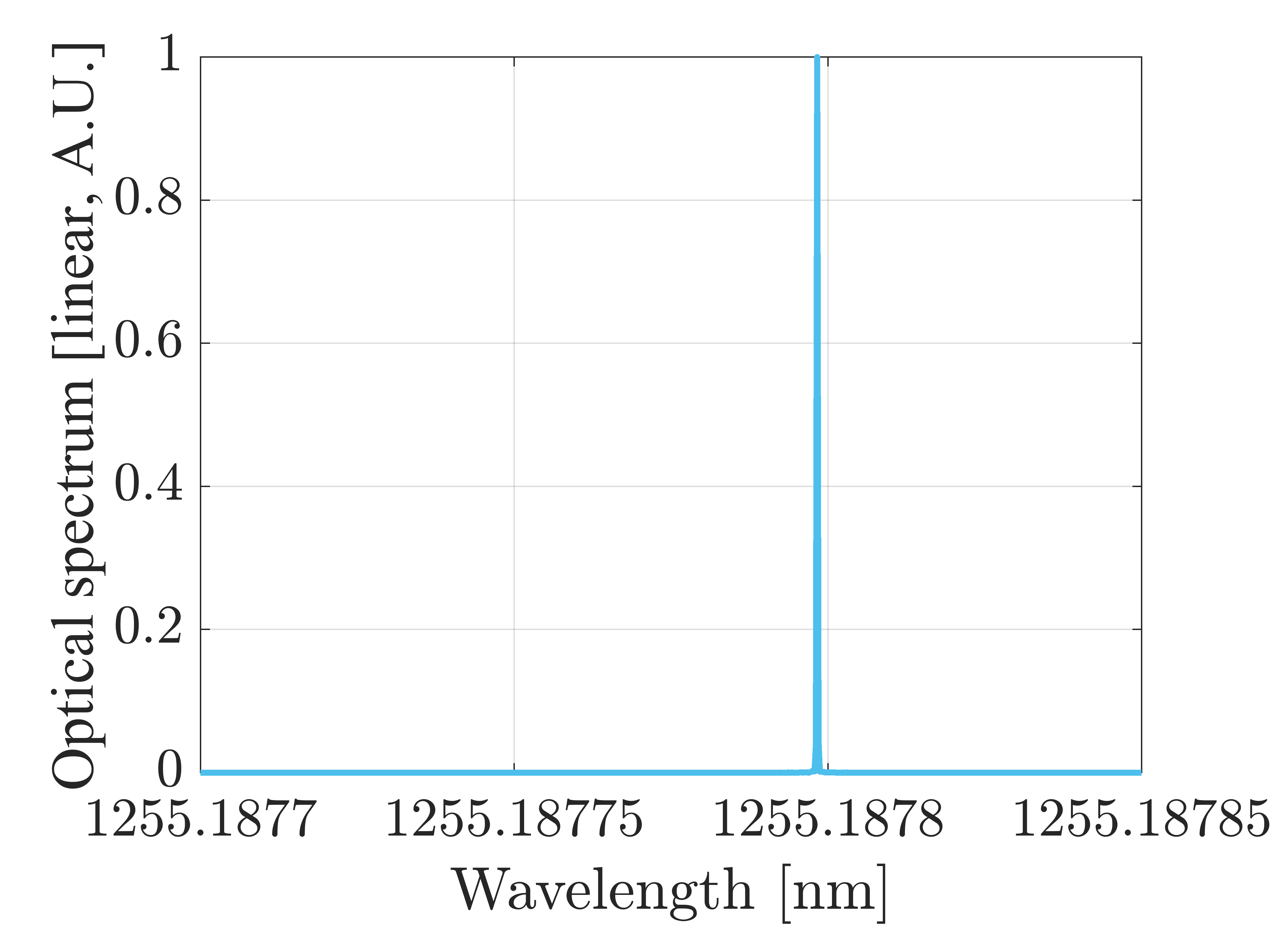}}
	\caption{Zoom of the optical spectra (normalized to their maximum) around the mode $q$ $=$ $0$
for $\Delta I$ $=$ \SI{40}{\milli\ampere} (violet line), $\Delta I$ $=$ \SI{60}{\milli\ampere} (green line), $\Delta I$ $=$ \SI{80}{\milli\ampere} (black line) and $\Delta I$ $=$ \SI{200}{\milli\ampere} (cyan line).
The main lasing line and the side bands generated by FWM are evident in the unlocked regimes ($\Delta I$ $<$ \SI{100}{\milli\ampere}) while it disappears when the modes are equally distant, i.e in the locked regime ($\Delta I$ $>$ \SI{200}{\milli\ampere}). }
	\label{fig5}
\end{figure}


\section {Discussion}

The numerical evidences reported in the previous Section regarding the phase-locking of the lasing lines might be interpreted by extending to several modes the results derived in \cite{Landais} for the study of self-pulsing  DBR semiconductor lasers (based on quantum well active medium) where three cavity longitudinal modes are locked. Whereas in that case the  three lasing lines were initially unevenly separated because of the DBR dispersion, in the case we are simulating this non-uniform separation derives from the dispersion of the QD active medium. The nonlinear interaction between the electric field and the gain medium (Four Wave Mixing) generates sub-bands close to each lasing line. These extra lines act as internally self-generated optical injection terms, pulling the instantaneous angular frequency of the main optical lines. 

For sufficiently high bias currents, when the intracavity field is strong enough, 
this optical injection process converges to the only possible equilibrium configuration, where all the lines are all equally spaced in frequency and exhibit a constant phase difference. This configuration corresponds to the OFC self-generation shown in Fig.\,\ref{fig4}.\\

On the contrary, in the low current regime, the power of the generated side band lines is not strong enough, and the FWM lines will not be able to force the locking: as a result, the system remains unlocked. In this condition, the high RIN of each single line in the low frequency region (reported for example as the red curve of Fig.\,\ref{fig3bis}a) is the consequence of the beating between the side-bands and the main lasing line. 

This behavior is clearly observable in Fig.\,\ref{fig5}, where the optical line $q=0$ are presented for injected currents ranging from \SI{40}{\milli\ampere} to \SI{200}{\milli\ampere} above threshold. For low values of the injected current (Fig.\,\ref{fig5}a) the FWM generated side lines around the main lasing peak are clearly visible, indicating that the modes power is not high enough for the FWM to induce a clean locking of the optical lines.
Increasing the current (and thus the power) the side lines amplitudes reduce (Fig.\,\ref{fig5}b,c) and completely disappear in the locked condition achieved (Fig.\,\ref{fig5}d). A similar self-mode-locking mechanism has been also introduced in \cite{Hugi2012, Villares2015} to interpret the comb formation in broad band in quantum cascade lasers . 


Finally, on a more fundamental level, we may identify the global indicators of the degree of coherence in the considered multi-wavelength QD laser, like $M_{\sigma_ {\delta P}}$ and $M_{\sigma\Delta\Phi}$ plotted in Fig.\,\ref{indi}, as "order" parameters for this complex physical system. 
Hence, their discontinuous transition across the boundary between the unlocked (``disordered'') and locked (``ordered'') regimes (``phases'') when bias current is varied is a manifestation of the Landau symmetry principle \cite{Landau}, according to which a phase transition involving a spontaneous symmetry breaking is associated to an order parameter which is non-analytical at the critical (transition) point.
In particular, in first order transitions the discontinuity affects the order parameter itself, as seems to occur in our case. This analogy with first order phase transition might also help explaining the coexistence of the locked and unlocked ``phases'' for a given interval of pump current (region B) that would play the role of an inverse temperature (\cite{Marconi,Lugiato}) and might indicate that the transition from unlocked to locked regimes occurs through a nucleation mechanism. 


\section{Conclusions}

We simulated the multi-mode dynamics of a single section Fabry-Perot QD laser using a TDTW approach that correctly describes the material gain and dispersion and takes into account for the carrier grating generated by the optical standing wave pattern formed in the laser cavity (SHB) and that is not washed out by diffusion as it happens in $1D$ or $2D$ semiconductor active regions. The SHB is at the origin of the single mode instability that leads to irregular multi-wavelength emission close to the lasing threshold (unlocked regime). For higher currents, degenerate and non degenerated FWM assure mode proliferation and phase-locking typical of OFC emission (locked regime). We characterized the degree of coherence in the system using both a spectrally resolved analysis through the estimation of the optical lines amplitude and the phase temporal fluctuations and more conventional, experimentally accessible, global quantifiers as the integrated RIN and the RF beat-note linewidth. Plotting these quantities {\it vs.} the pump current we were able to identify a value of bias current above which a quite sharp transition between the two regimes occurs and we observe low RIN and narrow inter-modal linewidth configurations associated with OFC self-generation. Our results are in good agreement with recently reported experimental evidences of self-mode-locking in QDash and QD single section lasers and add a physical insight into the understanding of these self-organization phenomena with undoubted impact on its applications in the field of e.g. high capacity optical communication.

\section{Appendix}
The coupled differential equations (\ref{fieldfasta})-(\ref{pop3fasta}) can be numerically solved using a finite differences scheme, discretizing both in time (with step $\Delta t$) and space (with step $\Delta z=v_g\Delta t$). 
In order to significantly reduce the computational time required by the numerical integration, it is possible to observe that the dynamics of the occupation probabilities $\rho_{0\,GS}$ and $\rho^+_{GS}$ is slower than the characteristic dephasing time $1/\Gamma$ of the corresponding inter-band transition \cite{Rossetti}. In this adiabatic limit, we obtain for the microscopic polarization the expression
$$
p^\pm_{k,j}={\rm j}\frac{d_{GS}}{\hbar\Gamma}\left[
(2\rho_{0\,GS\,k,j}-1)I_{k,j}+2\rho^\pm_{GS\,k,j}I^{\mp}_{k,j}
\right]
$$
where the terms
$$
I^\pm=\e^{-\Gamma\Delta t}I^\pm_{k,j-1}+\frac{\Gamma\Delta t}{2}\e^{-\Gamma \Delta t}E^\pm_{k,j-1}+\frac{\Gamma\Delta t}{2}E^\pm_{k,j}
$$
represent the  numerical approximation of the convolution integral between the field components and the material Lorentzian optical susceptibility and the subscripts $k$ and $j$ are the space and time steps counters, respectively.
The approximation $E^{\pm}_{k,j}$ of the forward and backward components of the electric fields $E^\pm(k\Delta z, j\Delta t)$ in Eq. (\ref{fieldfasta}) can be expressed as
\begin{equation}\label{eq:numericmodel1}
E^{\pm}_{k,j}=E^{\pm}_{k\mp 1,j-1}+\Delta z\left\{-\frac{\alpha_{wg}}{2}\frac{E^\pm_{k,j}+E^\pm_{k\mp 1,j-1}}{2}
-{\rm j}\frac{g_{0,GS}\hbar\Gamma}{d_{GS}}\frac{p^\pm_{k,j}+p^\pm_{k\mp 1,j-1}}{2}\right\}
\end{equation}
obtaining therefore
\begin{eqnarray}
    E^\pm_{k,j}&=&E^\pm_{k\mp 1,j-p}+\frac{\Delta z}{2}\left\{
    -\frac{\alpha_{wg}}{2}\left[E^\pm_{k,j}+E^\pm_{k\pm 1,j-1}\right]\right.\nonumber\\
    &\phantom{=}&+G_{0;k,j}\left[\e^{-\Gamma\Delta t}I^\pm_{k,j-1}+
    \frac{\Gamma\Delta t}{2}\e^{-\Gamma\Delta t}E^\pm_{k,j-1}+\frac{\Gamma\Delta t}{2}E^\pm_{k,j}
    \right]+G_{0;k\mp 1,j-1}I^\pm_{k\mp 1,j-1}\nonumber\\
    &\phantom{=}&+\left.G^\pm_{k,j}\left[\e^{-\Gamma\Delta t}I^\mp_{k,j-1}+
    \frac{\Gamma\Delta t}{2}\e^{-\Gamma\Delta t}E^\mp_{k,j-1}+\frac{\Gamma\Delta t}{2}E^\mp_{k,j}
    \right]+G^\pm_{k\mp 1,j-1} I^\mp_{k\mp 1,j-1}\right\}\label{eq:numericmodel2}\\
\end{eqnarray}
with $G_{0;k,j}=g_0(2\rho_{0;k,j}-1)$ and $G^\pm_{k,j}=2g_0(\rho_{k,j}^\pm)$;
for sake fo brevity the terms related to the spontaneous emission, calculated as in \cite{Rossetti}, are not reported in Eqs. (\ref{eq:numericmodel1}) and (\ref{eq:numericmodel2}).
 
Introducing the spatial and temporal dependent coefficients
$A_{0;k,j}=\Delta z\left(\Gamma\Delta tG_{0;k,j}-\alpha_{wg}\right)/4$,
    $A^+_{k,j}=\Delta z\Gamma\Delta tG^+_{k,j}/4$, $A^-_{k,j}=A^{+*}_{k,j}$
    and
\begin{eqnarray*}
    C^\pm_{(k\mp 1,k),(j,j-1)}&=&\frac{\Delta z}{2}\left\{
    -\frac{\alpha_{wg}}{2}E^\pm_{k\pm 1,j-1}
    +G_{0;k,j}\left[\e^{-\Gamma\Delta t}I^\pm_{k,j-1}+
    \frac{\Gamma\Delta t}{2}\e^{-\Gamma\Delta t}E^\pm_{k,j-1}
    \right]\right.\\\nonumber
&\phantom{=}&+G^\pm_{k,j}\left[\e^{-\Gamma\Delta t}I^\mp_{k,j-1}+
    \frac{\Gamma\Delta t}{2}\e^{-\Gamma\Delta t}E^\mp_{k,j-1}
    \right]\\
    &\phantom{=}&+G_{0;k\mp 1,j-1}I^\pm_{k\mp 1,j-1}
+\left.G^\pm_{k\mp 1,j-1} I^\mp_{k\mp 1,j-1}\phantom{\frac{1}{1}}\right\}
\end{eqnarray*}
we finally obtain the $2\times 2$ complex linear system
\begin{equation*}
    \left[ \begin{array}{cc}
        1-A_{0;k,j} & -A^+_{k,j} \\
        -A^{+*}_{k,j} & 1-A_{0;k,j}
    \end{array} \right]
    \left[ \begin{array}{c}
        E^+_{k,j}\\
        E^-_{k,j}
    \end{array} \right]
    =
    \left[ \begin{array}{c}
        C^+_{(k-1,k),(j,j-1)}+E^+_{k-1,j-1}\\
        C^-_{(k+1,k),(j,j-1)}+E^-_{k+1,j-1}
    \end{array} \right]
\end{equation*}
which has to be solved in each longitudinal slice to calculate the new approximation of the fields $E^\pm_{k,j}$. It has to be observed that the coupling between the forward and backward components of the field depends only on the coefficients $A^+_{k,j}$, related to the spatial carrier grating.
\section{Acknowledgments}
This work has been supported by the Fondazione CRT under the action ``La ricerca dei Talenti''. 

The Authors would like to acknowledge Prof. I. Montrosset for the fruitful discussions and helpful suggestions.
\bibliographystyle{osajnl}


\end{document}